\documentclass[fleqn,usenatbib]{mnras}
\usepackage{float}
\usepackage{newtxtext,newtxmath}
\usepackage{subcaption}

\usepackage[T1]{fontenc}

\DeclareRobustCommand{\VAN}[3]{#2}
\let\VANthebibliography\thebibliography
\def\thebibliography{\DeclareRobustCommand{\VAN}[3]{##3}\VANthebibliography}

\usepackage{xcolor}
\usepackage{graphicx}	
\usepackage{amsmath}	
\usepackage{lscape} 
\usepackage{pdflscape}

\defcitealias{2016Laigle}{L16}
\defcitealias{2018Jin}{J18}
\title[Obscured IR quasars in the COSMOS field]{A panchromatic view of infrared quasars: excess star formation and radio emission in the most heavily obscured systems}

\author[C. Andonie et al.]{Carolina Andonie,$^{1}$\thanks{E-mail: carolina.p.andonie@durham.ac.uk}
David M. Alexander,$^{1}$
David Rosario,$^{2, 1}$
Brivael Laloux,$^{1,3}$
Antonis Georgakakis,$^{3}$
\newauthor
Leah K. Morabito,$^{1, 4}$
Carolin Villforth,$^{5}$
Mathilda Avirett-Mackenzie,$^{5}$
Gabriela Calistro Rivera,$^{6}$
\newauthor
Agnese Del Moro,$^{7}$
Sotiria Fotopoulou,$^{8}$
Chris Harrison,$^{2}$
Andrea Lapi,$^{9}$
James Petley,$^{1}$
Grayson Petter,$^{10}$
\newauthor
and Francesco Shankar$^{11}$ \\
$^{1}$ Centre for Extragalactic Astronomy, Department of Physics, Durham University, Durham, DH1 3LE, UK\\
$^{2}$ School of Mathematics, Statistics and Physics, Newcastle University, Newcastle upon Tyne, NE1 7RU, UK\\
$^{3}$ Institute for Astronomy \& Astrophysics, National Observatory of Athens, V. Paulou \& I. Metaxa 11532, Athens, Greece\\
$^{4}$ Institute for Computational Cosmology, Department of Physics, University of Durham, South Road, Durham DH1 3LE, UK\\
$^{5}$ Department of Physics, University of Bath, Claverton Down, Bath BA2 7AY, UK \\
$^{6}$ European Southern Observatory (ESO), Karl-Schwarzschild-Straße 2, 85748 Garching bei München, Germany\\
$^{7}$ German Aerospace Center (DLR), Space Operation and Astronaut Training, Oberpfaffenhofen, D-82234 Weßling, Germany \\
$^{8}$ HH Wills Physics Laboratory, University of Bristol, Tyndall Avenue, Bristol BS8 1TL, UK \\
$^{9}$ SISSA, Via Bonomea 265, I-34136 Trieste, Italy\\
$^{10}$ Department of Physics and Astronomy, Dartmouth College, 6127 Wilder Laboratory, Hanover, NH 03755, USA\\
$^{11}$ School of Physics \& Astronomy, University of Southampton, Highfield, Southampton SO17 1BJ, UK\\
}

\date{Accepted XXX. Received YYY; in original form ZZZ}

\pubyear{2015}

\begin{document}
\label{firstpage}
\pagerange{\pageref{firstpage}--\pageref{lastpage}}

\maketitle


\begin{abstract}

\noindent To understand the Active Galactic Nuclei (AGN) phenomenon and their impact on the evolution of galaxies, a complete AGN census is required; however, finding heavily obscured AGNs is observationally challenging. Here we use the deep and extensive multi-wavelength data in the COSMOS field to select a complete sample of 578 infrared (IR) quasars ($L_{\rm AGN,IR}>10^{45}\rm \: erg\: s^{-1}$) at $z<3$, with minimal obscuration bias, using detailed UV-to-far IR spectral energy distribution (SED) fitting. We complement our SED constraints with X-ray and radio observations to further investigate the properties of the sample. Overall, 322 of the IR quasars are detected by {\it Chandra} and have individual X-ray spectral constraints. From a combination of X-ray stacking and $L_{\rm 2-10\rm keV}$ -- $L_{\rm 6\: \mu m}$ analyses, we show that the majority of the X-ray faint and undetected quasars are heavily obscured (many are likely Compton thick), highlighting the effectiveness of the mid-IR band to find obscured AGNs. We find that 355 ($\approx$~61\%) IR quasars are obscured ($N_{\rm H}>10^{22}\rm \: cm^{-2}$) and identify differences in the average properties between the obscured and unobscured quasars: (1) obscured quasars have star-formation rates $\approx 3$ times higher than unobscured systems for no significant difference in stellar mass and (2) obscured quasars have stronger radio emission than unobscured systems, with a radio-loudness parameter $\approx 0.2 \rm \: dex$ higher. These results are inconsistent with a simple orientation model but in general agreement with either extreme host-galaxy obscuration towards the obscured quasars or a scenario where obscured quasars are an early phase in the evolution of quasars.

\end{abstract}




\section{Introduction}

It is widely known that all massive galaxies host a supermassive black hole (SMBH) at their centre \citep[e.g.,][]{1998Magorrian, 2003Marconi}, implying that all massive galaxies must have hosted an active galactic nuclei (AGN) during their lifetime \citep[e.g.,][]{1984Rees}. A complete census of AGN activity is therefore crucial to constrain how these SMBH and their galaxies grew and to, consequently, estimate the radiative efficiency of the BH growth \citep[e.g.,][]{2010Shankar,2020Shankar,2012Alexander}. The identification of all AGNs may also reveal dependencies on the presence of obscuration with the environment (from small to large scales) since obscured AGNs are more likely to reside in more dust-rich and gas-rich environments than unobscured AGNs \citep[e.g.,][]{2012Alexander,2018HickoxyAlexander}. 

A complete census of AGNs is also crucial to understand the role of SMBH accretion in galaxy evolution. The tight correlation between the SMBH mass and the stellar mass of the galaxy bulge \citep[e.g.,][]{2000Ferrarese,2000Gebhardt,2003Marconi} provides evidence for a connection between the growth of SMBHs and their host galaxies. First-order evidence for this connected growth is seen in the evolution of the star formation rate (SFR) and SMBH accretion rate over cosmic time, which both show a peak at redshift $z \approx 2$ and then start to decline towards the present day \citep[e.g.,][]{1998Boyle,2010Aird,2015ABrandt}. However, it is still uncertain how this apparent evolution between galaxies and SMBH is related.

It is now well established that the majority of the AGNs population is obscured \citep[e.g.,][]{2014Ueda, 2015Aird, 2015Buchner}. An efficient method to identify obscured AGNs is X-ray observations since (1) X-ray emission appears to be a near universal property of AGNs, (2) X-ray photons can penetrate high column densities of gas, and (3) the contamination from the host galaxy (i.e., due to supernova remnants or X-ray binaries) is negligible in most cases \citep[e.g.,][]{2015ABrandt}. Indeed, the signature of the emission from the entire AGN population including even the most extreme Compton-thick (CT, $\rm N_H \geq 1.5 \times 10^{24}\: cm^{-2}$) systems is, in principle, imprinted on the cosmic X-ray background (CXB) \citep[e.g,.][]{1995Comastri, 2004Bauer, 2007Gilli, 2014Ueda, 2019Ananna}. However, even the deepest X-ray surveys struggle to identify the most obscured AGNs when the column density is CT since most of the X-ray emission is absorbed. 

An alternative approach to identifying obscured AGNs is MIR emission. The mid-infrared (MIR) emission from AGNs originates after the absorption and re-emission of UV light by circumnuclear dust; consequently, MIR emission is even less biased against obscuration than X-ray observations. The drawback of the MIR band is that it can be highly contaminated by star formation (SF) emission from the host galaxy; therefore, separating the AGN from the host galaxy emission is not trivial. A powerful tool that allows for the separation of the AGN from the host galaxy emission, and the identification of luminous AGN even in strong star-forming galaxies is broad-band UV--far-IR Spectral Energy Distribution (SED) template fitting \citep[e.g.,][]{2013Berta,2013DelMoro,2014Rovilos, 2016DelMoro, 2016Calistro,2019Battisti,2021Kokorev}. SED decomposition is also an excellent technique for identifying obscured AGN with moderate levels of dust attenuation \citep[e.g.,][]{2014Rovilos, 2016DelMoro, 2021CalistroG} that have been missed even in the deepest X-ray surveys. 

With the aim of selecting a complete and unbiased set of obscured AGNs and investigating possible dependencies between AGN and host galaxy properties, we use the deep and extensive multi-wavelength data in the Cosmic Evolution Survey (COSMOS; \citealp{2007Scoville} ). One of the motivations of this work is to extend the previous research in our group on red and blue quasars to more obscured systems, where we explored the panchromatic properties of red quasars using Bayesian SED fitting \citep[][]{2021CalistroG}, and it was found that red quasars have excess radio emission compared to blue quasars, due to compact ($<2\rm \: kpc$) radio sources \citep[][]{2019Klindt, 2020Fawcett, 2020Rosario, 2021Rosario, fawcett2022}. To achieve this objective, we perform a detailed analysis of the optical/UV to far IR (FIR) SED of the sources in COSMOS to identify a complete set of IR-emitting quasars, including both obscured and unobscured systems. Our paper is organized as follows. In Section\,\ref{sec:data}, we describe the multi-wavelength data used in our analyses, and in Section\,\ref{sec:sedsample}, we present our SED fitting approach and IR-quasar sample selection. In Section\,\ref{sec:results}, we then investigate the multi-wavelength properties of the IR quasar sample and compare several key properties of the obscured and unobscured systems, carefully controlling for redshift and AGN luminosity. 
More specifically, in Section\,
\ref{sec:res:comp} we estimate the completeness of our sample using X-ray observations and compare to other IR-selection approaches; in Section\, \ref{sec:res:SEDs} we investigate the overall UV/optical--far-IR SEDs of the obscured and unobscured quasars; in Section\,\ref{sec:res:X-ray}, we use published X-ray spectral fitting results of the \textit{Chandra} \textit{COSMOS-Legacy} survey \citep{2016Civano} to study the X-ray properties of the quasars; in Sections\, \ref{sec:res:SFR} and \ref{sec:res:Mstar}, we present the SFRs and stellar masses of the sample derived from our SED analyses; and in Section\,\ref{sec:res:radio} we use 1.4 and 3 GHz VLA data to investigate the radio properties of the sample. We finally discuss and put in context our results in Section\,\ref{sec:discussion} and present our conclusions in Section\,\ref{sec:conclusions}.

Throughout this work, we adopt a concordance cosmology \citep{2013Hinshaw}, and a Chabrier initial mass function \citep[][]{2003Chabrier}.

\section{Data} \label{sec:data}

In this section, we describe the multiwavelength datasets used to select our IR quasar sample. 

\subsection{Multiwavelength photometry}

To construct the SEDs of our sources, we use the public COSMOS2015 (\citealp{2016Laigle}, \citetalias{2016Laigle}) and \citet{2018Jin} (\citetalias{2018Jin}) catalogues. \citetalias{2018Jin} used a "super-deblended" approach \citep{2018Liu} to derive IR and radio photometry for 194,428 galaxies in the COSMOS field, covering ground-based K-band, \textit{Spitzer} \citep[][]{2004Werner}  IRAC and MIPS, \textit{Herschel} PACS/SPIRE  \citep[][]{2010Pilbratt}, SCUBA2, AzTEC, MAMBO and VLA bands. K-band and radio source positions were used as priors for the deblending process, and were obtained from \citetalias{2016Laigle} and \cite{2013Muzzin}, respectively. We take the optical/UV and near-IR (NIR) photometry from the COSMOS2015 catalogue and the MIR and FIR photometry from \citetalias{2018Jin}. Overall, we include 20 photometric bands in the 0.1-500 $\mu \rm m$ range: the u$^*$ filter from CFHT; the V, B, r+, i+, and z++ filters from the Subaru Suprime-Cam \citep{2002Miyazaki}; the Y, J, H, and Ks filters from the UltraVISTA-DR2 survey \citep{2012McCracken}; \textit{Spitzer} IRAC fluxes at 3.6, 4.5, 5.8, and 8$\mu$m; \textit{Spitzer} MIPS at 24 $\mu$m; \textit{Herschel} PACS at 100 and 160 $\mu$m; and \textit{Herschel} SPIRE at 250, 350 and 500 $\mu$m. 

To assign the redshifts of our sources, we cross-match the positions of the \citetalias{2018Jin} catalogue with the quasar catalogue of SDSS DR14 \citep{2020Rakshit}, the compilation of spectroscopic redshifts of Salvato in. (prep), and the catalogue of optical and IR counterparts of the {\it Chandra COSMOS-Legacy} survey \citep{2016MarchesiA}, adopting a matching radius of 1$^{\prime \prime}$. We use the redshifts in order of priority: (1) \citet{2020Rakshit} (100 objects), (2) Salvato in. (prep) (3364 objects), (3) \citet{2016MarchesiA} (2699 objects), (4) spectroscopic redshifts from \citetalias{2018Jin} (23461 objects), (5) photometric redshifts from \citetalias{2016Laigle} (125464 objects). Overall, we have 26837 spectroscopic redshifts and 168270 photometric redshifts.

\subsection{Radio observations} \label{sec:data:radio}

To analyze the radio emission of our sample, we use as a first option the VLA-COSMOS 3 GHz survey \citep{2017Smolcic}, which is the largest and deepest radio survey in COSMOS. For this, we crossmatch the positions of the \citetalias{2018Jin} catalogue with the positions in \citet{2017Smolcic}, allowing 1$^{\prime \prime}$ of separation. Otherwise, we use the 1.4 GHz VLA fluxes reported by \citetalias{2018Jin}. We re-scale the 3 GHz fluxes from \citet{2017Smolcic} to 1.4 GHz fluxes, assuming that the radio emission can be describe as a powerlaw with an uniform spectral slope, $ S_{\nu} \propto \nu^{\alpha}$, with $\alpha=-0.9$. The adopted spectral slope correspond to the median calculated from the 1.4 GHz fluxes from the VLA-COSMOS \citep{2010Schinnerer} and the 3 GHz fluxes from \citet{2017Smolcic}. We did not use the 1.4 GHz fluxes from the VLA-COSMOS Survey since it is almost 10 times fainter than the 3 GHz survey; hence, it provides fluxes for just a small fraction of our sample. 



\subsection{X-ray data} \label{sec:Xraydata}

\subsubsection{The \textit{Chandra COSMOS-Legacy} survey}

The properties of the X-ray sources in the COSMOS field have been extensively studied in previous works. The deepest and most complete X-ray catalogue available is the \textit{Chandra COSMOS-Legacy} survey \citep[][]{2016Civano}, a 4.6 Ms \textit{Chandra} program that has observed the 2.2 deg$^2$ area of the COSMOS field, with a homogeneous exposure time of $\sim$ 160 ks, yielding a 2-10 keV flux limit of $\rm F_{2-10keV} \sim 1.9\times 10^{-15}  \: erg \: cm^{-2} \: s^{-1}$. \citet{2016MarchesiB} studied all sources in the \textit{COSMOS-Legacy} catalogue with more than 30 net counts in the 0.5-7 keV band and computed X-ray properties such as column density ($N_{\rm H}$), intrinsic 2-10 keV X-ray luminosity ($L_{2-10\rm keV}$) and photon index ($\Gamma$) for a final sample of 1949 sources. To calculate these properties, they initially fitted the X-ray spectra with an absorbed power-law model, allowing $N_{\rm H}$ and $\Gamma$ to vary (for sources with $<70$ spectral net counts, $\Gamma$ was set to a fixed value of 1.9). For sources obscured by $N_{\rm H}>10^{22} \rm \: cm^{-2}$, they added a second power-law component to account for the scattered component (e.g., scattered radiation without being absorbed by dust and gas). This approach performs well when the level of obscuration is modest ($N_{\rm H} \leq 10^{23} \rm \: cm^{-2}$). In a later study, \citet{2018Lanzuisi} performed a detailed analysis of 67 CT AGN candidates from \citet{2016MarchesiB}, using a physically motivated model that accounts for the photoelectric absorption, Compton scattering, reflection, and fluorescent emission lines (see Section 2.2 of  \citealp[][]{2018Lanzuisi} for details). Therefore, we use \citet{2018Lanzuisi} results for the 67 CT AGN candidates and \citet{2016MarchesiB} for the rest of the sources, following the recommendation of the main authors of both papers. The X-ray spectral properties used in this paper were acquired by private communication with the authors.

As a test, we also performed our analyses using the X-ray spectral fitting results in the COSMOS field from Laloux et al. (submitted), which adopts a Bayesian approach to constrain the AGN physical properties. We found small quantitative changes with respect to our study here and our overall conclusions do not qualitatively change.

To identify the X-ray counterparts of our sample, we cross-match the positions of \citetalias{2018Jin} with the optical and infrared positions of the X-ray sources in the COSMOS field from \citet{2016MarchesiA}, using a 1$^{\prime \prime}$ matching radius. In order of priority, we use (1) the K-band Ultravista positions, (2) the optical positions, and (3) the IRAC positions. We identified X-ray counterparts for 3437 sources.

\subsubsection{X-ray flux upper limits estimation} \label{sec:Xuplims}

For the sources that are not associated with an X-ray counterpart, we determine 3$\sigma$ upper limits to their X-ray flux. This calculation is based on aperture photometry and follows the methods described in \cite{Ruiz2022}. In brief, the X-ray photons in the 0.5-2\,keV and 2-7\,keV spectral bands are extracted at the source position within apertures of variable radius that correspond to the 70\% Encircled Energy Fraction (EEF) of the Chandra Point Spread Function. Background maps of 0.5-2\,keV and 2-7\,keV are then used to determine the background expectation value within the extraction apertures. The average exposure times within the same regions are calculated using the Chandra exposure maps.  It can be shown \citep[e.g.][]{Kraft1991, Ruiz2022} that the flux upper limit, $UL$, of a source within the aperture can be estimated by numerically solving the equation  

\begin{equation}\label{eq:age:CL}
C\cdot \int_{0}^{UL} P(f_X | N) \,df_X = C \dot L,
\end{equation}

\noindent where $C$ is a normalization constant, $N$ is the total number of photons within the extraction aperture, $P(f_X | N)$ is the Poisson probability of a source with flux $f_X$ given the observed number of photons, and $CL$ is the confidence level of the upper limit. Our analysis assumes $CL=99.7$\%, meaning that the true flux lies below the respective upper limit. This probability corresponds to the two-sided $3\sigma$ limit of a Gaussian distribution. The Poisson probability for a source with flux $f_X$ is

\begin{equation}
P(f_X | N)  = \frac{e^{-\lambda}\cdot \lambda^N}{N!},
\end{equation}

\noindent where $\lambda$ is the expected number of photons in the extraction cell for a source with flux $f_X$

\begin{equation}
\lambda = f_X \cdot EEF \cdot ECF\cdot t + B,
\end{equation}

\noindent where $t$ is the exposure time, $B$ is the background level, $ECF$ is the energy flux to photon flux conversion factor and depends on the spectral shape of the source. Given $N$, $B$, $t$ and $CL$, Equation\,\ref{eq:age:CL} can be numerically solved \citep[see][]{Ruiz2022} to infer the flux upper limit. For the X-ray spectral shape, we adopt a power-law with $\Gamma=1.8$. The inferred X-ray flux upper limit depends only mildly on this choice. 
 
\section{SED fitting and sample selection} \label{sec:sedsample}

In this section, we describe the models we use to fit the SED of the sources and the process to select the final IR quasar sample. 

\subsection{SED modeling} \label{sec:sedfitting}
 
We fit the rest-frame $0.1-500 \: \mu \rm m$ SED of the objects using the multicomponent Bayesian SED fitting package \textsc{fortesfit} \footnote{https://github.com/vikalibrate/FortesFit} \citep[][]{2019Rosario}. In fitting the SEDs, we consider the unabsorbed stellar emission modelled by \citet{2003BC}; the UV emission from the accretion disk following the model of \citet{2006Richards}; the IR emission from the torus, based on the empirical DECOMPIR AGN model of \citet{2011Mullaney}; and a semi-empirical star-forming galaxy model which reproduces the full range of dust temperatures observed in galaxies \citep{2014Dale}. 


The stellar emission template has as free parameters the age and mass of the stellar population, and the star-formation timescale. As the emission can be affected by galactic extinction, we attenuate the templates according to the Milky Way reddening law from \citet{2000Calzetti}, adding the reddening $\rm E(B-V)_{SP}$ as a free parameter. Similarly, the accretion disk emission might also be altered by dust along the line of sight, therefore we apply the \citet{1984Prevot} reddening law for the Small Magellanic Clouds, which has been shown to be representative of Type-1 AGN \citep[e.g.,][]{2004Hopkins,2009Salvato}, and describe well the SEDs of red and blue QSOs \citep[][]{2021CalistroG, fawcett2022}. Hence, the free parameters of the accretion disk (AD) model are the luminosity at 2500{\AA} ($L_{2500}$) and the reddening $\rm E(B-V)_{AD}$. The free parameters of the star-forming galaxy model are the integrated $8-1000 \:\mu \rm m$ galaxy luminosity from the SF ($L_{\rm SF, IR}$) and a shape parameter that describes a wide range of of PAH emission and spectral shapes for normal star-forming galaxies. Finally, the torus template combines a smoothly broken powerlaw and a black body. The parameters of the broken power-law component are the short-wavelength slope ($\Gamma_{s}$), the wavelength of the break ($\lambda_{Brk}$), and the long-wavelength slope ($\Gamma_{L}$), and the parameters of the black body component are the peak wavelength of the black body ($\lambda_{BB}$), and the temperature of the short end ($T_{BB, s}$). Following \citet[][]{2011Mullaney}, we fix $\lambda_{Brk} = 19 \rm \: \mu m $ and $\lambda_{BB}=40 \rm \: \mu m$. 

We note that in this work, we use a simple one-shape parameter template to model the star formation emission in FIR wavelengths, as we are only interested in constraining the SF luminosity of our sample. To test how our fits change with other star formation templates, we fit the $2.2-500\rm \: \mu m$ SED of our IR quasar candidates (see Section\,\ref{sec:qsoselection} for details), modelling the star formation emission with the \citet{2014Draine} templates, which are parameterized by the polycyclic
aromatic hydrocarbon mass fraction, the lower cutoff of the starlight intensity distribution, and the fraction of the dust heated by starlight. We find that the SF luminosities are entirely consistent between the \citet{2014Draine} and \citet{2014Dale} models, indicating that although the templates adopted in this work might be comparatively simple, they are also comprehensive, and successfully constrain the broad range of contributions from star formation to the IR emission.

In summary, we fit 20 photometric bands in the rest-frame $0.1-500\mu \rm m$ range with a model with 12 free parameters. Table\,\ref{t:params} lists the parameters of each model, their description, and the value ranges considered to perform the fits. 


\begin{table*}
\centering

 \begin{tabular}{lcccc} 
 \hline
 \hline
\noalign{\smallskip}

\bf Component &\bf  Parameter & \bf Description &{\bf Range} & {\bf Reference} \\
\noalign{\smallskip}
 \hline
\noalign{\smallskip}
\noalign{\smallskip}
Stellar population emission (SP)  & $\log M_{\star} \rm \: [M_{\odot}]$ & stellar mass & $\rm [6,13]$ & \citet[][]{2003BC}\\
 \noalign{\smallskip}
 & $\log t'\rm \: [Gyr]$  & age of the stellar population & $\rm [8,10.1]$&\\
 \noalign{\smallskip}
 & $\log \tau\rm \: [log\:Gyr]$  & star-formation timescale  & $\rm [0.01,15]$ &\\
 \noalign{\smallskip}
& $\rm E(B-V)_{SP} \: [mag]$  & galaxy reddening & $\rm [0,0.5]$ & \citet[][]{2000Calzetti}\\
\noalign{\medskip}

Accretion disk (AD) & ${\rm log \:} L_{2500}\rm \: [erg \: s^{-1} \: Hz^{-1}]$ & luminosity at 2500{\AA} & $[26,36]$ & \citet[][]{2006Richards} \\
 \noalign{\smallskip}
 & $\rm E(B-V)_{AD} \: [mag]$  & accretion disk reddening & $\rm [0,1]$ & \citet[][]{1984Prevot}\\
\noalign{\medskip}

Torus & ${\rm log \:} L_{\rm AGN, IR} \rm \: [erg \: s^{-1}]$ & $8-1000 \mu m$ AGN luminosity & $[38,48]$ & \citet[][]{2011Mullaney} \\
 \noalign{\smallskip}
& $\Gamma_{\rm s}$  & short-wavelength slope & $[-0.3,2.1]$ & \\
 \noalign{\smallskip}
& $\Gamma_{\rm L}$  & long-wavelength slope & $[-1,0.5]$ &\\
 \noalign{\smallskip}
 & $T_{\rm BB, s} \rm \: [K]$  & hot dust temperature & $[800,1900]$ \\
 \noalign{\medskip} 

Star formation (SF) & ${\rm log \:} L_{\rm SF, IR} \rm \: [erg \: s^{-1}]$ & $8-1000 \rm \: \mu m$ star-formation luminosity & $[42,48]$ &\citet[][]{2014Dale}\\
 \noalign{\smallskip}
& $\alpha_{\rm SF}$  & shape parameter & $[0.06,4]$ &\\
 \noalign{\smallskip}
\hline
 \noalign{\smallskip}
 \end{tabular}
 \caption{Description and allowed ranges of the free parameters for the components in our SED fitting models, including references for more details of each model. } \label{t:params}
\end{table*}

\subsection{IR quasars selection} \label{sec:qsoselection}

\begin{figure}
\centering
\includegraphics[trim={0 9cm 0 0},clip,scale=0.25]{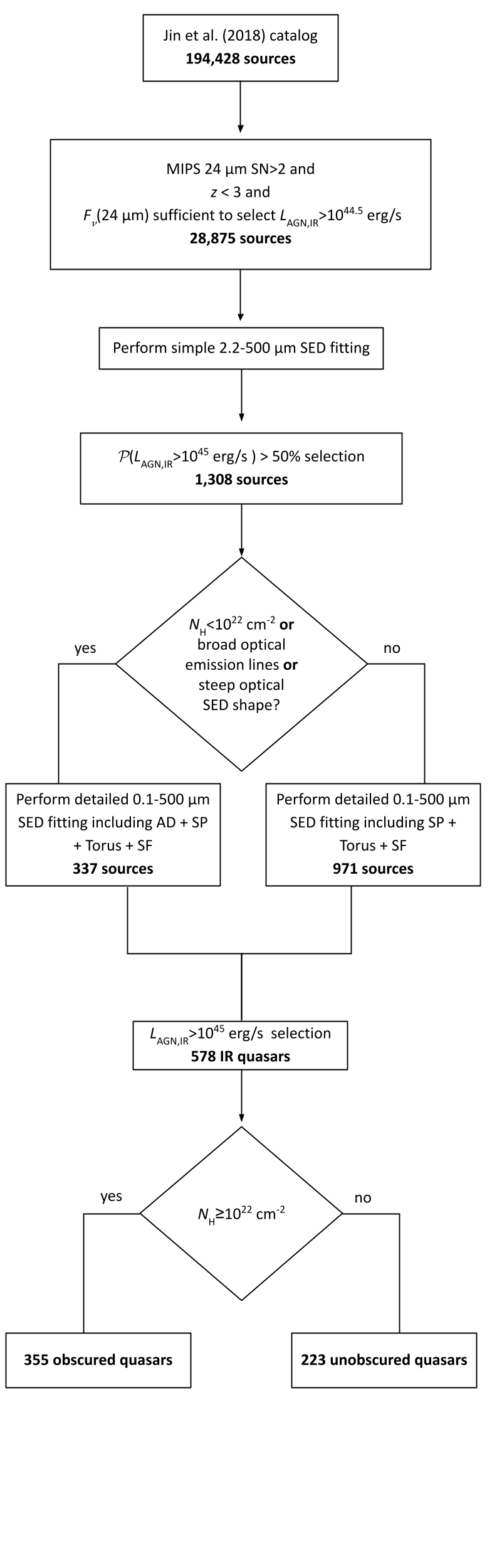}
\caption{A schematic representation of our IR quasar sample selection process. We started selecting the sources in the \citet{2018Jin} catalogue with a 2$\sigma$ detection at $24 \:\mu \rm m$ and $z<3$. We then apply a cut in $F_{\nu} (24 \: \mu \rm m)$ flux to select the sources that would have $8-1000\rm \: \mu m$ luminosity $L_{\rm AGN,IR}>10^{44.5}\rm \: erg\: s^{-1}$ if we assume that all of the $F_{\nu} (24 \: \mu \rm m)$ flux comes from the AGN. We fit the 2.2-500 $\mu \rm m$ SED of those sources with a simple model (see Section\,\ref{sec:qsoselection} for details) and select all sources with a probability higher than 50$\%$ of having $L_{\rm AGN,IR}>10^{45}\rm \: erg\: s^{-1}$; we refit the rest-frame 0.1-500 $\mu \rm m$ SEDs of these IR quasar candidates with a more complex model, including the accretion disk when necessary. We finally select the sources with $L_{\rm AGN,IR}>10^{45}\rm \: erg\: s^{-1}$, and then split the sample into the obscured and unobscured IR quasars, using $N_{\rm H} = 10^{22}\rm \: cm^{-2}$ as the unobscured/obscured threshold (see Section\,\ref{sec:res:X-ray} for details). }

\label{fig:selection}
\end{figure}

The selection process of our sample has several steps and is outlined in Figure\,\ref{fig:selection}. The process consists of two main components: 1) we first preselect candidate IR quasars based on 24 $\mu \rm m$ detections and flux and then fit the SEDs of these objects with a simple model to identify the most likely luminous AGN in the \citetalias{2018Jin} catalogue; 2) we then refit these sources with a more detailed AGN model and select the IR quasars with  $L_{\rm AGN,IR}>10^{45}\rm \: erg\: s^{-1}$. In our analyses, we only consider the sources with $\rm SNR> 2 $ at 24 $\mu \rm m$, as it is the main band that samples the rest-frame MIR, and enables us to break some of the degeneracy of SF versus AGN emission in that important diagnostic wavelength regime. We only select sources with redshift $z<3$, because at $z\gtrsim 3$, the observed 24 $\mu \rm m$ band no longer constrains the AGN emission, as it is equivalent to $\lesssim 6 \: \mu \rm m$ rest-frame. Since we are only interested in quasars with $L_{\rm AGN,IR}>10^{45}\rm \: erg\: s^{-1}$, we apply a cut in the $F_{\nu} (24\mu \rm m)$ flux to select the sources that would have $L_{\rm AGN,IR}>10^{44.5}\rm \: erg\: s^{-1}$, conservatively assuming that all of the $F_{\nu} (24\mu \rm m)$ flux is produced by the AGN. We achieve this by computing the $F_{\nu} (24\mu \rm m)$ curve as a function of redshift using the DECOMPIR AGN template, for $\Gamma_{L}=0.2$, $T_{BB, s}=1500\rm K$, and $\Gamma_{s}=0.8$, which gives a lower limit for $F_{\nu} (24\mu \rm m)$. The application of these cuts results in 28,875 IR quasar candidates. 

We fit the SEDs for all IR quasar candidates and obtain the posterior distribution for each parameter as outputs. To save computational time, we pre-select the IR quasars by only fitting the $2.2-500 \:\mu \rm m$ SED; hence, we do not consider the optical photometry, and so do not include the AD in our modelling, and we let free to vary only the short-wavelength slope of the torus component. The rest of the parameters of the torus component are fixed to $\Gamma_{L}=0.2$ and $T_{BB, s}=1500\rm K$, which are the values of the mean intrinsic SED found by \citet{2011Mullaney}. We compute the $L_{\rm AGN,IR}$ cumulative distribution function of each object to estimate the probability of each object of having an AGN luminosity $L_{\rm AGN,IR}>10^{45}\rm \: erg\: s^{-1}$. We find 1308 sources with a $\geq 50\%$ probability of hosting an AGN component in the quasar regime. We then incorporate the optical photometry to fit the rest-frame $0.1-500 \:\mu \rm m$ SED of the IR quasar candidates, considering a version of the DECOMPIR model that allows all three torus parameters to vary across the prescribed ranges listed in Table\,\ref{t:params}. This time, we include the AD when the source is either X-ray unobscured (i.e., $N_{\rm H} < 10^{22} \: \rm cm^{-2}$; see Section\,\ref{sec:res:X-ray} for details), has broad lines (BL) in the optical spectra reported by \citet{2013Rosario}, or has a steep UV-optical SED, suggesting the presence of an accretion disk component. For the X-ray obscured sources ($N_{\rm H} \geq 10^{22} \: \rm cm^{-2}$), we associate all optical emission to the stellar population since the emission from the AD will be extinguished.

It is difficult to constrain the stellar population and accretion disk emissions of the unobscured sources because both components overlap in the optical band. Therefore, we need to apply a prior to the stellar mass, $M_{\star}$, and the AD luminosity, $L_{2500}$, to improve the measurements of these parameters. There are 142 IR quasar candidates with broad lines and an estimation of their SMBH mass ($M_{\rm BH}$) in \citet{2013Rosario}. For those objects, we use as a prior the $M_{\star}$ estimated from $M_{\rm BH}$--$M_{\star}$ relation reported in \citet{2010Merloni}. For the sources without a $M_{\rm BH}$ estimation, we use as a prior the mean $M_{\star}$ of the obscured sources ($M_{\rm \star,obs}/\rm  M_{\odot} = 10.71$ and $\sigma = 0.3 \rm \: dex $), which we compute with \textsc{PosteriorStacker} (see Section\,\ref{sec:uplims}). For the AD luminosity, we use the well-established $L_{2500}$--$L_{\rm 2keV}$ relation for quasars reported by \citet{2016Lusso} to calculate a prior for $L_{2500}$, when an object has an X-ray luminosity measurement. In both cases, we apply a Gaussian prior centered in the estimated value for $M_{\star}$ or $L_{2500}$ with a standard deviation of $\sigma = 0.3 \rm \: dex $. 

\begin{figure}
\centering
\includegraphics[scale=0.42]{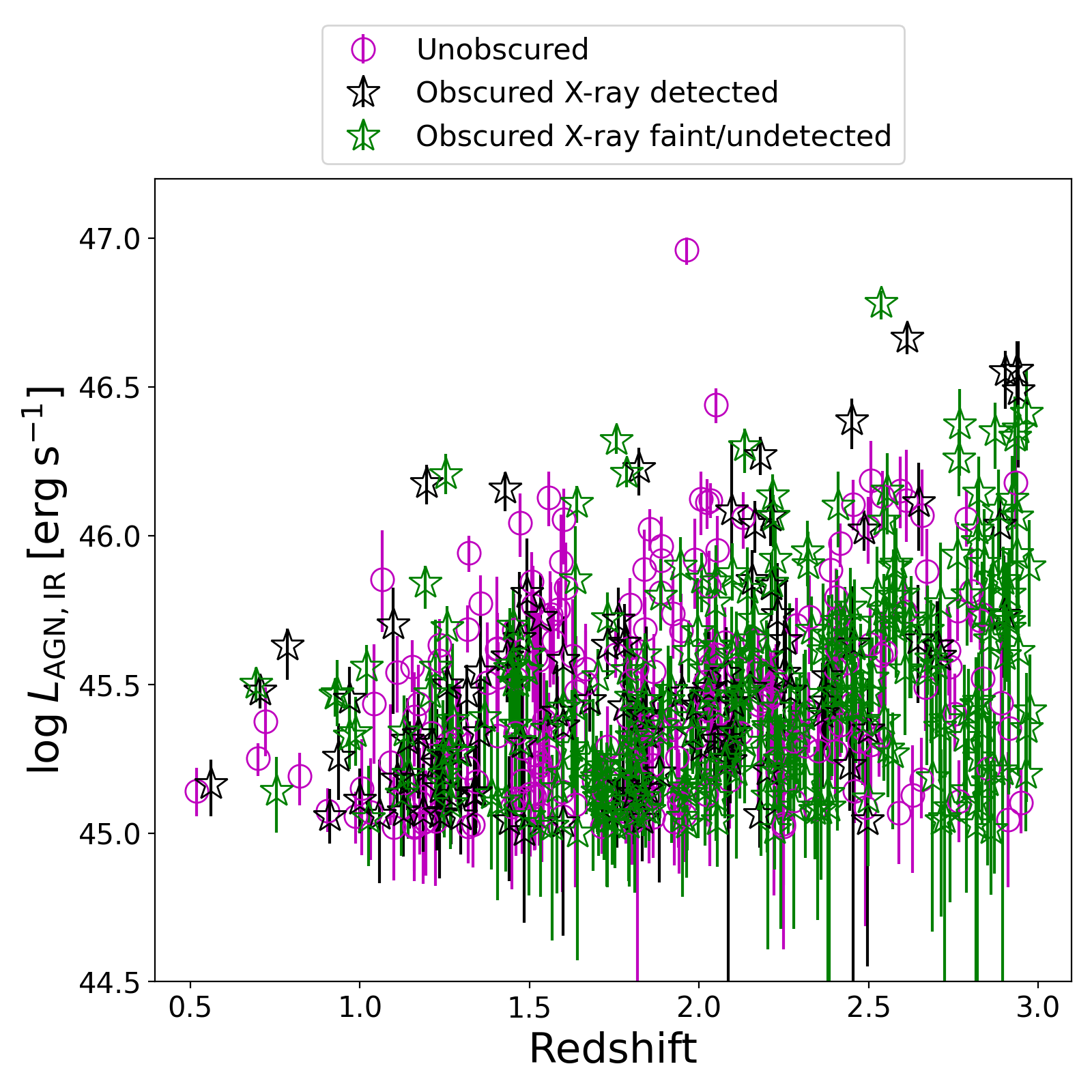}
\caption{$L_{\rm AGN,IR}$ (the 8-1000 $\mu \rm m$ AGN luminosity) against the redshift of our IR quasar sample. The error bars represent 1$\sigma$ uncertainties. Magenta open circles are objects with $N_{\rm H}<10^{22} \: \rm cm^{-2}$ from X-ray spectral fitting, black open stars are objects with $N_{\rm H}\geq 10^{22} \: \rm cm^{-2}$ from X-ray spectral fitting, and green open stars are objects undetected or faint (i.e., insufficient counts to perform X-ray spectral fitting) in the X-ray band but which have properties implying $N_{\rm H}\geq 10^{22} \: \rm cm^{-2}$ (see Section\,\ref{sec:res:X-ray} for details).}

\label{fig:LAGN_z}
\end{figure}

\begin{figure*}
\centering
\includegraphics[scale=0.38]{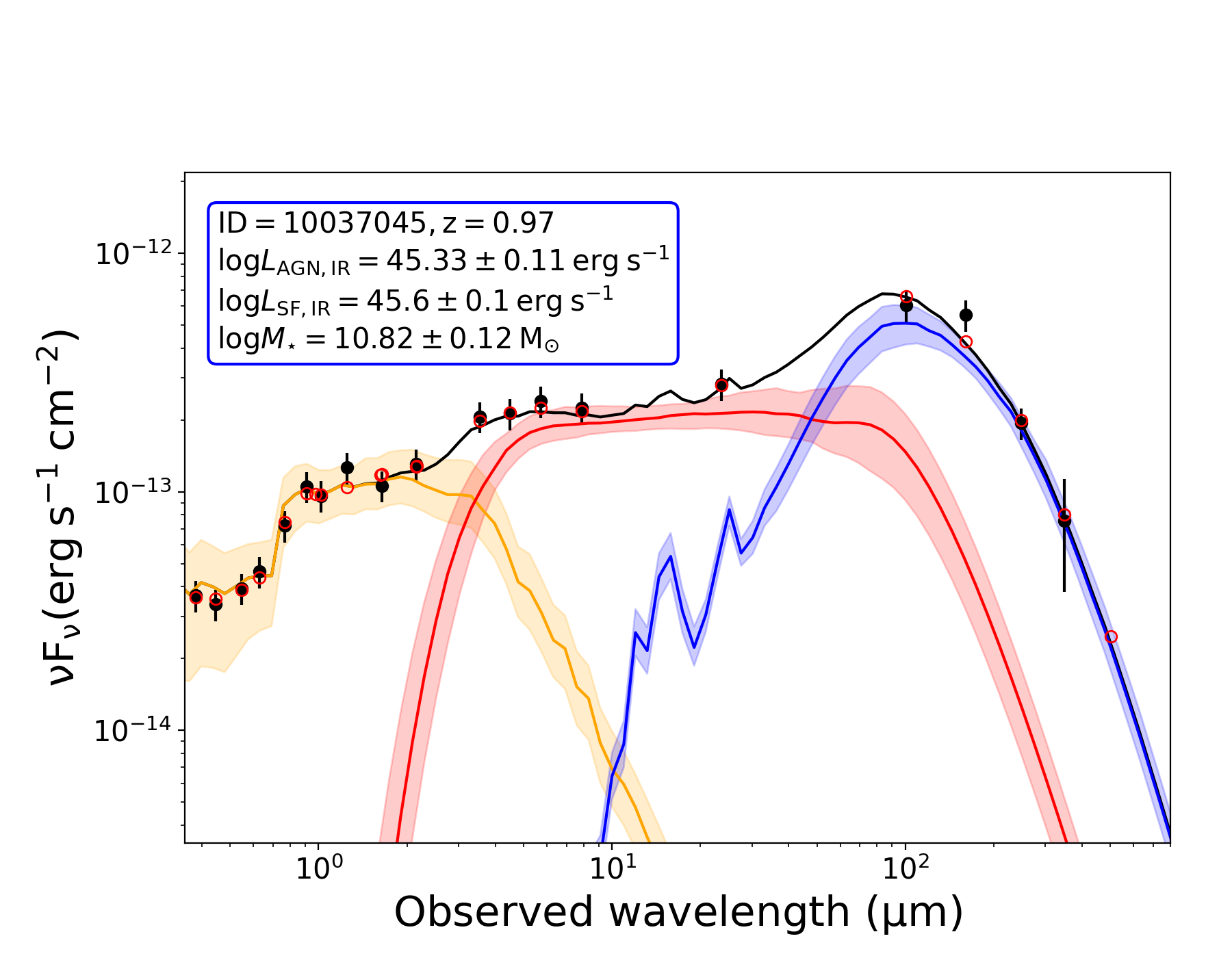}
\includegraphics[scale=0.38]{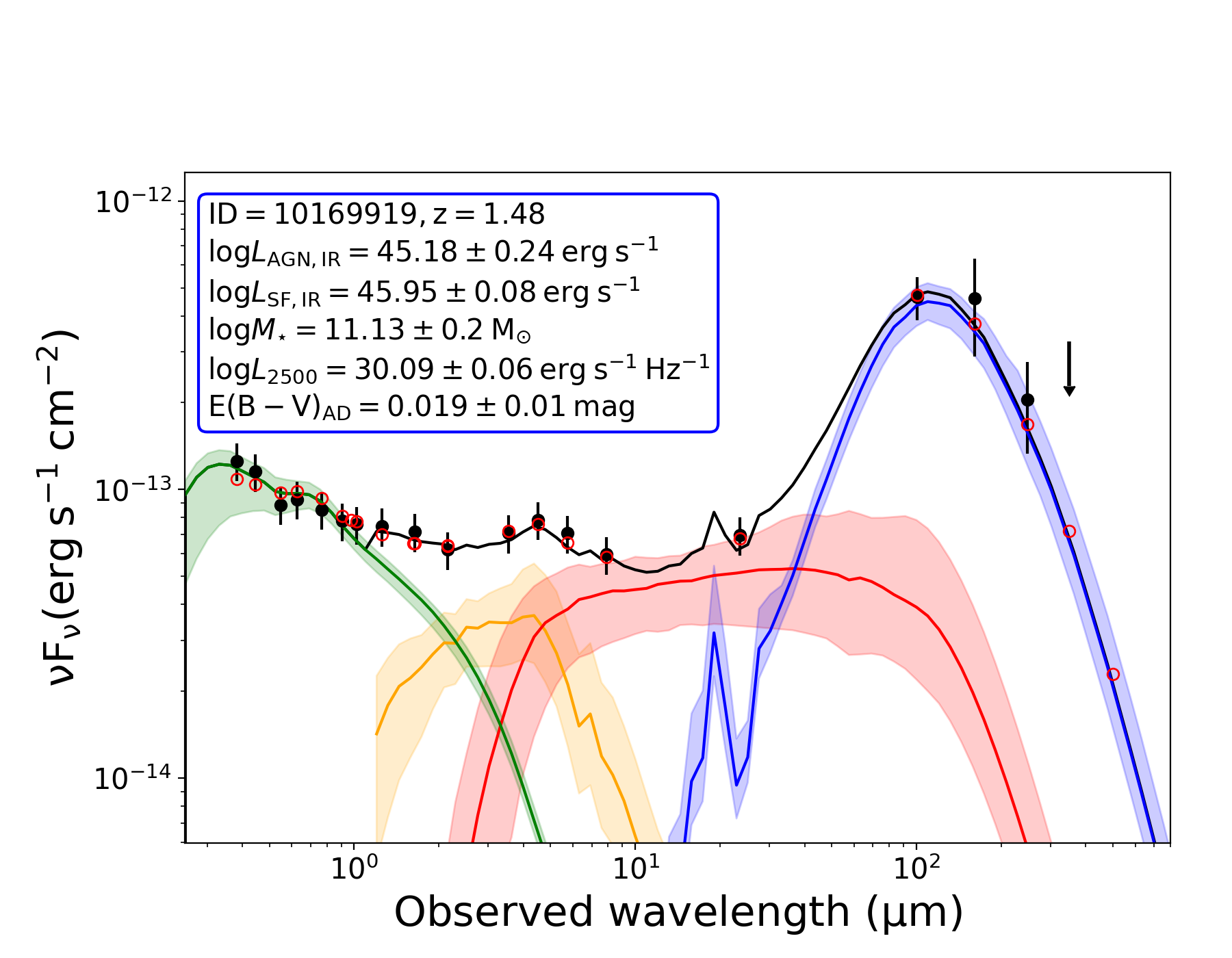}

\caption{Example best-fitting SED models for an obscured quasar (left panel) and an unobscured quasar (right panel). Black points with error bars are the photometric data with their respective 1$\sigma$ uncertainties, and red empty circles are the best-fitting model photometry. The blue curves represent the dust-obscured SF component, the red curves are the torus components, the orange curves are the stellar population emission, and the green curves are the accretion disk emission. The spread in the curves represents the approximate $1 \sigma$ scatter in the SED components as constrained by \textsc{fortesfit}. The black line is the sum of all the best-fitting components. Key derived parameters and 1~$\sigma$ uncertainties from the best-fitting SED models are shown in the blue outlined box. These figures show samples of the full SED and are only meant to guide the reader; hence, they should not be used to measure the merit of the fit. }
\label{fig:SEDs_ex}
\end{figure*}

The final IR quasar sample comprises 578 sources with $L_{\rm AGN,IR}>10^{45}\rm \: erg\: s^{-1}$. Although we select sources down to $\rm SNR>2$ in MIPS $24\: \rm \mu m$, we note that the vast majority (561/578, 97\%) of the IR quasars have a $\rm SNR>5$, indicating that our sample should not be impacted by unreliable or blended $24\: \rm \mu m$ photometry. 
Figure\,\ref{fig:LAGN_z} shows the $L_{\rm AGN,IR}$ against the redshift of our sample. The redshift of the IR quasars ranges between 0.5 and 3. We find that the majority of our IR quasars have $10^{45}<L_{\rm AGN,IR}/{\rm \: erg\: s^{-1}}<10^{46}$, but there are 56 very powerful AGNs with $L_{\rm AGN,IR}>10^{46}\rm \: erg\: s^{-1}$. The bolometric luminosities of our quasars span the range $L_{\rm AGN, bol}=[10^{45.5}, 10^{47.4}]{\rm \: erg\: s^{-1}}$; these are calculated as $L_{\rm AGN, bol} \approx 3 \times L_{\rm AGN,IR}$, assuming that $L_{\rm AGN, bol} \approx 8 \times L_{\rm AGN, 6\mu m}$ \citep{2006Richards}, and that the median ratio $L_{\rm AGN, 6\mu m}/ L_{\rm AGN,IR} \approx 0.36$ for our quasars. Overall, 393 ($\approx$~68\%) IR quasars are detected in the X-ray band: 322 sources are bright enough to have X-ray spectral constraints (see Section\,\ref{sec:res:X-ray} for details) and 71 sources are faint (i.e., not enough counts to perform X-ray spectral fitting). The remaining 185 sources are undetected in the X-ray band. Furthermore, 321 quasars have spectroscopic redshifts and 257 have photometric redshifts. Figure\,\ref{fig:SEDs_ex} shows examples of the SED fitting results for two objects, an obscured (left panel) and an unobscured (right panel) IR quasar. Table\,\ref{A:bfpar} reports the best-fitting SED parameters of our IR quasars sample.


\subsection{Upper limits and mean values estimation} \label{sec:uplims}

In order to decide if a parameter is well constrained, we take two different approaches depending on the case. For the AGN and SF luminosity, we analyse the 1st percentile of the posterior distributions. In both cases, the distributions have two peaks. For $L_{\rm AGN,IR}$, the peaks are at $\log \: L_{\rm AGN,IR, 1p}/[\rm erg/s]<38.5$ and $\log \: L_{\rm AGN,IR, 1p}/[\rm erg/s] \gtrsim 44.5$, while for $L_{\rm SF,IR, 1p}$, the peaks are at $\log L_{\rm SF,IR, 1p}/[\rm erg/s]<42.5$ and $\log L_{\rm SF,IR, 1p}/[\rm erg/s]\gtrsim 45$. Therefore, we consider that the AGN and SF components of the SED are well constrained when $\log L_{\rm AGN,IR, 1p}/[\rm erg/s] \geq 38.5$ and $\log L_{\rm SF,IR, 1p}/[\rm erg/s] \geq 42.5$, respectively.


In the case of the stellar masses of the unobscured quasars, we use a Gaussian prior to constrain the parameter; hence, we cannot take the same approach as when using a uniform prior, as is the case for the AGN and SF luminosities. This time, we use the Kullback-Leibler Divergence (KLD) parameter, which is commonly used for model comparison in Bayesian frameworks and estimates the distance between two distributions. In our case, those distributions are the Gaussian prior and the stellar mass posterior distribution. A low KLD parameter indicates that the prior and the posterior distributions are very similar and that during the SED fitting process the model does not learn anything from the data and just reproduces the prior. On the other hand, a high KLD value indicates that the model learns from the data and the prior helps to constrain the parameter better. We investigate the distribution of the stellar mass KLD parameter and find that it has two peaks: at ${\rm KLD} < 0.2$ and ${\rm KLD} \gtrsim 0.5$. Therefore, we consider that the stellar masses are well constrained when ${\rm KLD} \geq 0.2$. 

In order to account for the upper limits and unconstrained parameters, to compute the median values of the SF luminosity and stellar masses we use Posterior Stacker \footnote{\url{https://github.com/JohannesBuchner/PosteriorStacker}}, a code that takes as input samples of the posterior distributions of a sample of objects and models them using a Gaussian model \citep{2020Baronchelli}.

\section{Results} \label{sec:results}

This section presents the multi-wavelength properties of the IR quasar sample and explores potential differences between the obscured and unobscured IR quasars. We assess the completeness of our IR SED selection approach in Section\,\ref{sec:res:comp}, explore the different variety of SED shapes we obtain in Section\,\ref{sec:res:SEDs}, and present the X-ray properties of our IR quasar sample in Section\,\ref{sec:res:X-ray}. We then compare the SF properties (Section\,\ref{sec:res:SFR}), the stellar mass (Section\,\ref{sec:res:Mstar}), and the radio emission (Section\,\ref{sec:res:radio}) between the obscured and unobscured IR quasars. 

\subsection{IR quasar sample completeness} \label{sec:res:comp}

To investigate whether our IR SED fitting misses any luminous AGN, we fit the rest-frame 2.2-500 $\mu \rm m$ SED of all X-ray sources in the \textit{Chandra} COSMOS-Legacy survey at $z<3$, following the procedure described in Section\,\ref{sec:qsoselection}.  While X-ray surveys will miss the most heavily obscured AGN (as demonstrated in Section~\ref{sec:res:X-ray}), they are effective at identifying less obscured AGN down to modest X-ray luminosities, even in systems with bright host-galaxy emission where we would expect our IR-AGN identification approach to struggle.

\begin{figure}
\centering
\includegraphics[scale=0.42]{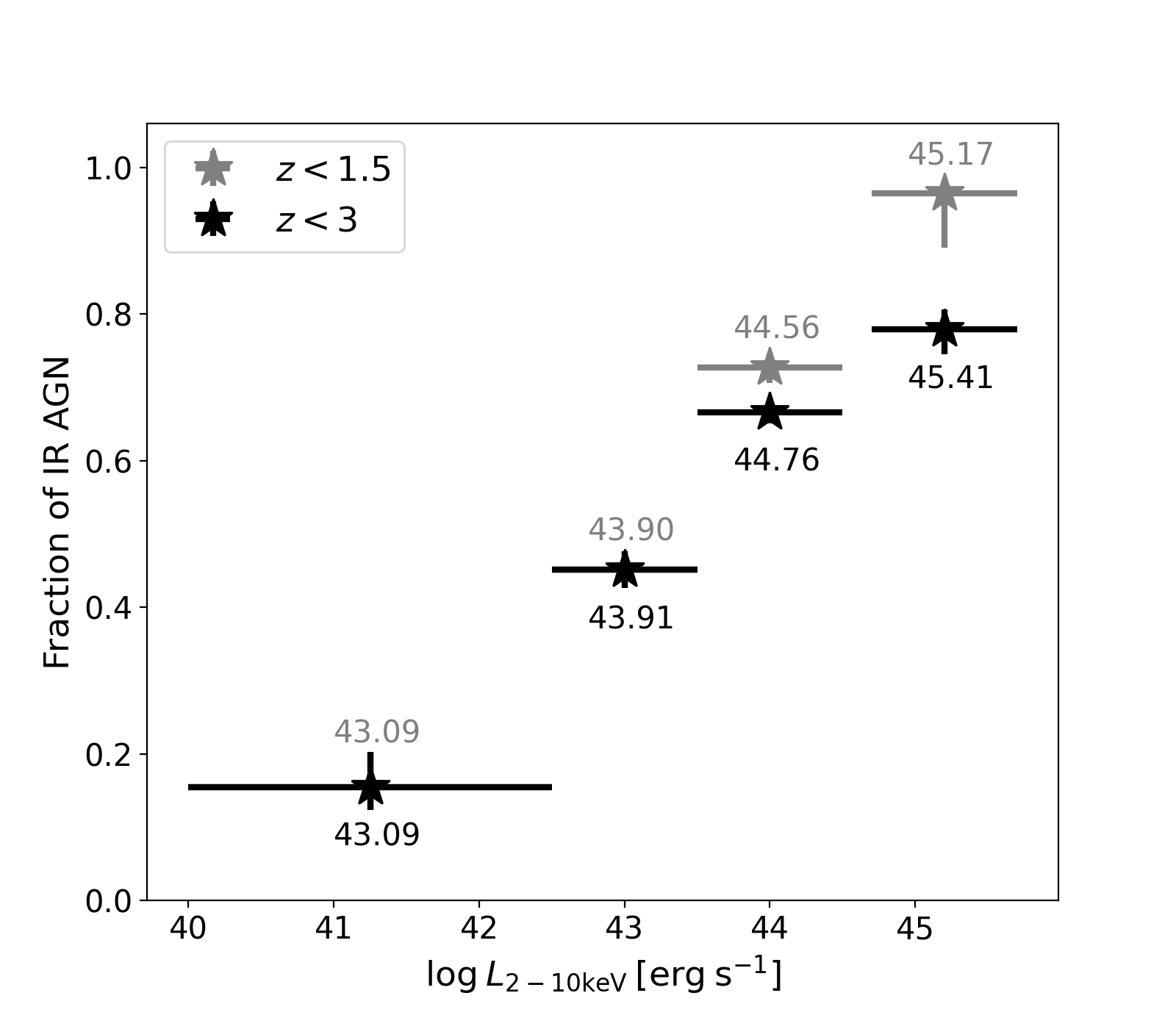}
\caption{Fraction of X-ray AGNs that are identified as IR AGNs as a function of the intrinsic X-ray luminosity ($L_{2-10 \rm \: keV}$) at $z<1.5$ (grey stars) and $z<3$ (black stars) for the sources in the \textit{Chandra} COSMOS Legacy survey. The error bars in the x-axes represent the size of the luminosity bins and the y-axes indicate the binomial errors. The black and grey text reports the mean SED best-fitting $8-1000\rm \: \mu m$ AGN luminosity in $\rm \log(erg\:s^{-1})$ for each X-ray luminosity and redshift bin.  }

\label{fig:IRAGN_Lx}
\end{figure}

Figure\,\ref{fig:IRAGN_Lx} shows the fraction of X-ray sources with a well constrained AGN component in the IR SED fits as a function of the X-ray luminosity. Overall, our SED fitting identifies $60\pm 1\%$ of the X-ray AGNs; however, the fraction of IR AGN increases with X-ray luminosity, as expected given lower-luminosity AGN are challenging to identify in SED-fitting approaches due to an increasing contamination from the host-galaxy emission which dilutes the signature of the AGN. In the highest-luminosity bin, equivalent to our quasar IR-luminosity threshold ($L_{2-10 \rm \: keV}>10^{44.5} \rm \: erg \: s^{-1}$), we reach a maximum completeness of $96^{+0.1}_{-0.7} \%$ at $z<1.5$ (1/28 missed) and $78\pm 0.3 \%$ at $z<3$ (40/181 missed).
Seventeen of the 40 missed AGNs have a 99th upper-limit percentile of $L_{\rm AGN,IR}>10^{45} \rm \: erg \: s^{-1}$, which means that the SED fitting does not identify an AGN, but we cannot rule out the presence of an IR quasar in these systems, typically due to lack of any {\it Herschel} detections. The 99th upper-limit percentile on the AGN luminosity is $L_{\rm AGN,IR}<10^{45} \rm \: erg \: s^{-1}$ for the other 23 sources: 5 lack a 2$\sigma$ detection at 24$\mu \rm m$, which is fundamental to constrain the AGN component in the IR SED, 12 have an SED fully dominated by the host galaxy, and 6 have both an unconstrained AGN and SF component due to lack a of photometric detections.

As another comparison, we also select AGNs using different well-established MIR colour-colour selections: we find that the \citet[][]{2005Stern} {\it Spitzer}-IRAC wedge selects $49\pm 1\%$, \citet[][]{2012Donley} selects $30\pm 1\%$ (also {\it Spitzer}-IRAC photometry), and the \citet[][]{2018Assef} 90\% reliability selects $11\pm 0.8\%$ of the X-ray AGNs using just {\it WISE} photometry. We note that our SED fitting approach is not only more complete but also more reliable than any MIR colour selection since it considers the redshift of the source and separate the AGN from the SF emission in the IR band. Therefore, our method might not find all of the X-ray AGNs, but it is cleaner and more complete than a typical MIR colour selection technique. For example, if we apply these MIR colour-colour selections to our IR quasar sample, then find that \citet[][]{2005Stern} would miss 13\%, \citet[][]{2012Donley} would miss 20\%, and \citet[][]{2018Assef} would miss 40\% of our IR quasars.

\subsection{The UV/optical-to-FIR SEDs of IR quasars} \label{sec:res:SEDs}

\begin{figure*}
\centering
\includegraphics[scale=0.55,trim={0 0 3cm 0},clip]{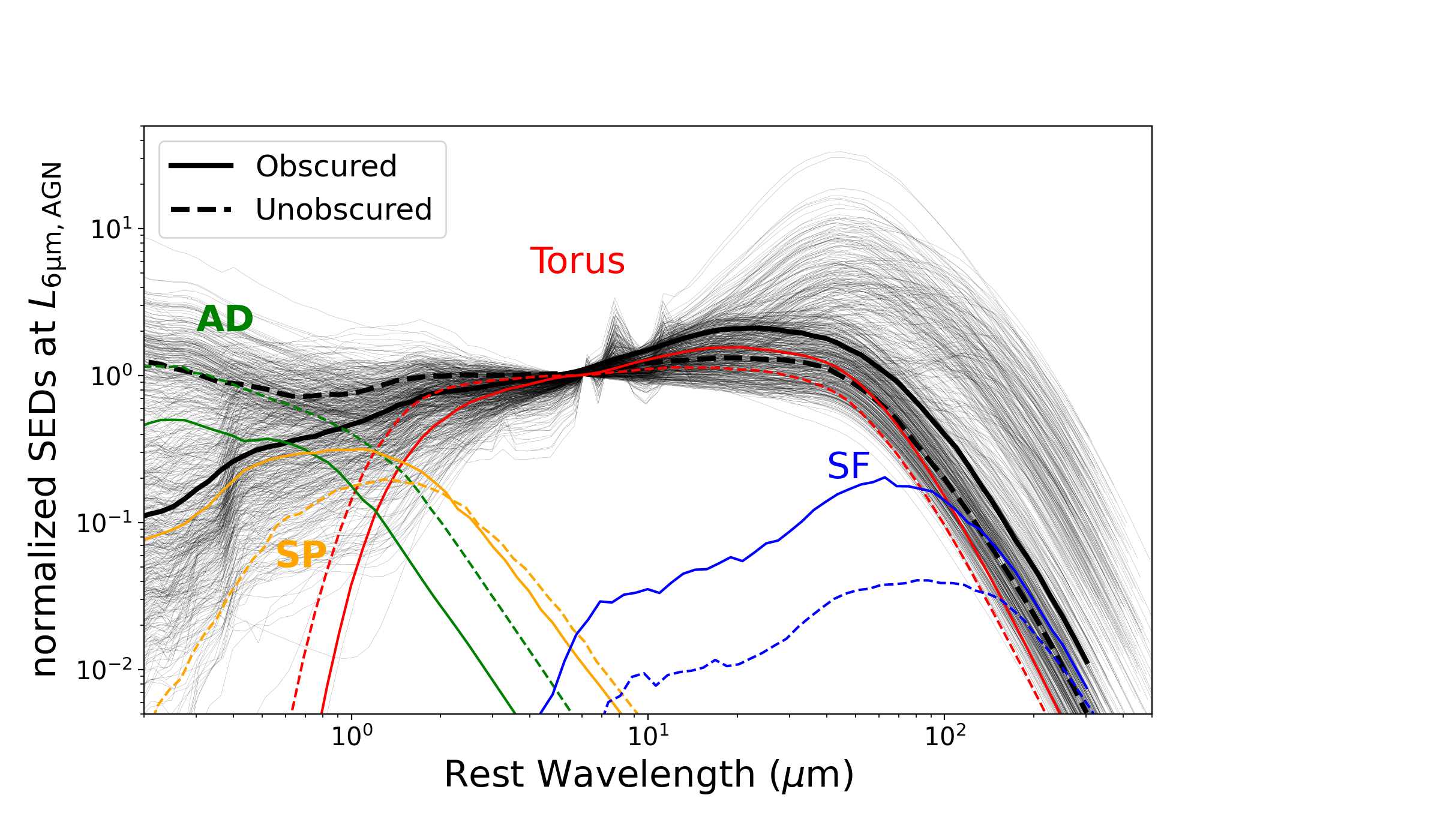}
\caption{Composite and individual best-fitting SED models for the IR quasar sample. Grey thin curves are the best-fitting SED models for each quasar, normalized at rest-frame $6\rm \: \mu m$ luminosity. Thick continuous and dotted curves represent the median SED for the obscured ($N_{\rm H} \geq 10^{22}\: \rm cm^{-2}$) and unobscured ($N_{\rm H} < 10^{22} \: \rm cm^{-2}$) quasars, respectively. The figure also shows the median SED of each component: the accretion disk (green), the stellar population emission (orange), the torus (red), and the dust-obscured SF (blue). The median AD SED of the obscured quasars was calculated only for the sources that required an AD model. The median SEDs are normalized to the median AGN luminosity at $6\rm \: \mu m$ in order to obtain the relative contribution of each component with respect to the AGN emission.   } 

\label{fig:composite_SEDs}
\end{figure*}

We present the best-fitting SED models of our IR quasar sample in Figure\,\ref{fig:composite_SEDs}. The SEDs are normalized by the total AGN luminosity at rest-frame $6 \: \rm \mu m$. The figure also plots the overall median SEDs of the obscured and unobscured quasars, together with the relative contribution of each fitted component to the total emission. We find a wide variety of SED shapes, and identify clear differences in the median SEDs between the unobscured and obscured quasars. The unobscured quasars are brighter in the optical band due to the AD emission but have a weaker SF component relative to the AGN emission. Additionally, the AGN torus component of the unobscured quasars is hotter than that of obscured quasars, showing stronger MIR emission and weaker FIR emission. In the following sections, we perform a detailed analysis of the properties of the IR quasars obtained from the SED fits to understand the observed differences in the SED shapes.

\subsection{X-ray properties} \label{sec:res:X-ray}

In our sample, 322 of the 578 ($56\%$) quasars are bright enough ($>30$ net counts, hereafter the X-ray bright sample) in the 0.5-7 keV band to have X-ray spectral constraints \citep[][]{2016MarchesiB, 2018Lanzuisi}, while 71 sources are X-ray detected but faint (i.e., $<30$ net counts), without X-ray spectral fitting results. The remaining 185 sources (32\%) are X-ray undetected. 

\begin{figure*}
\centering
\includegraphics[scale=0.38]{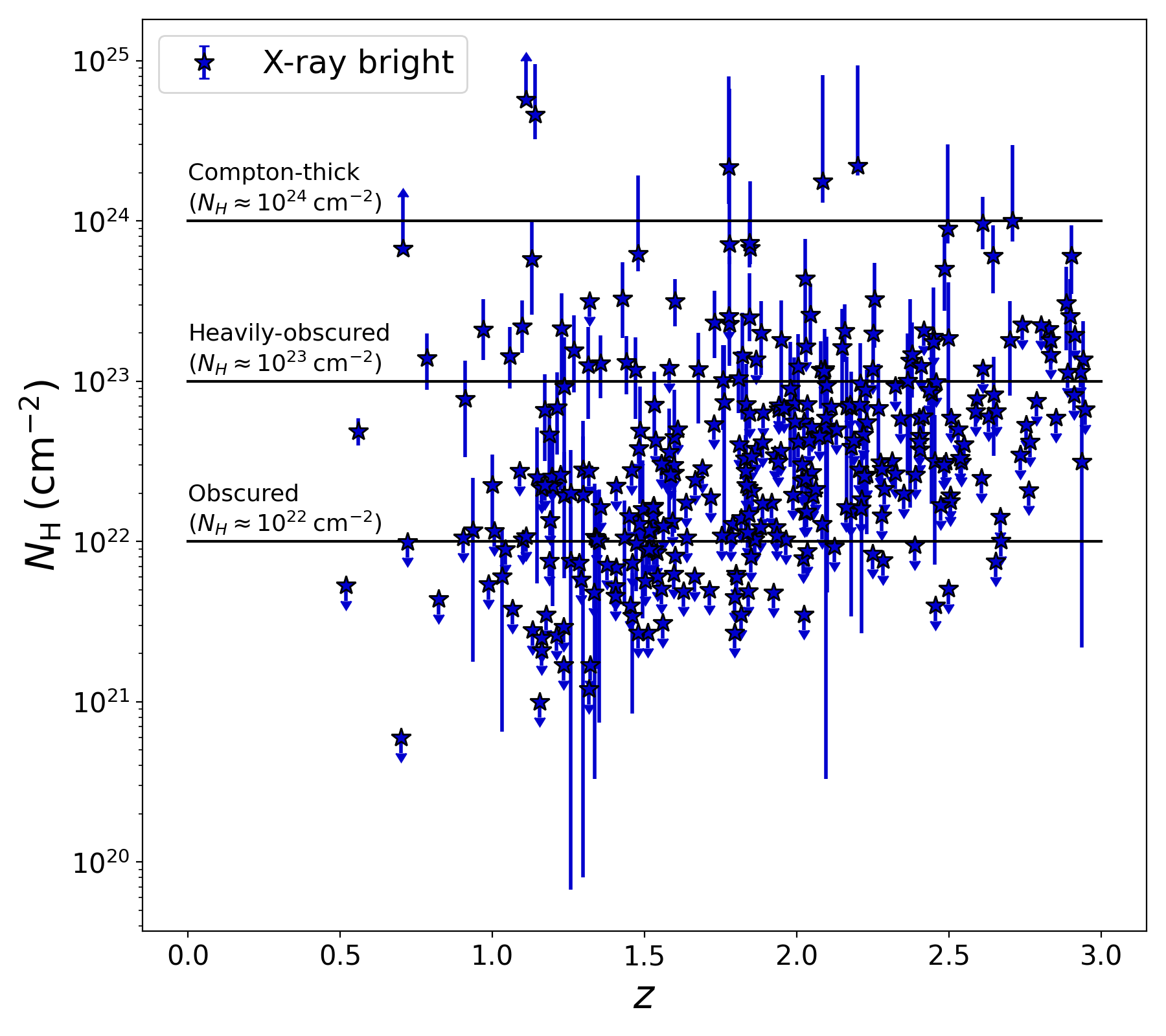}
\includegraphics[scale=0.38]{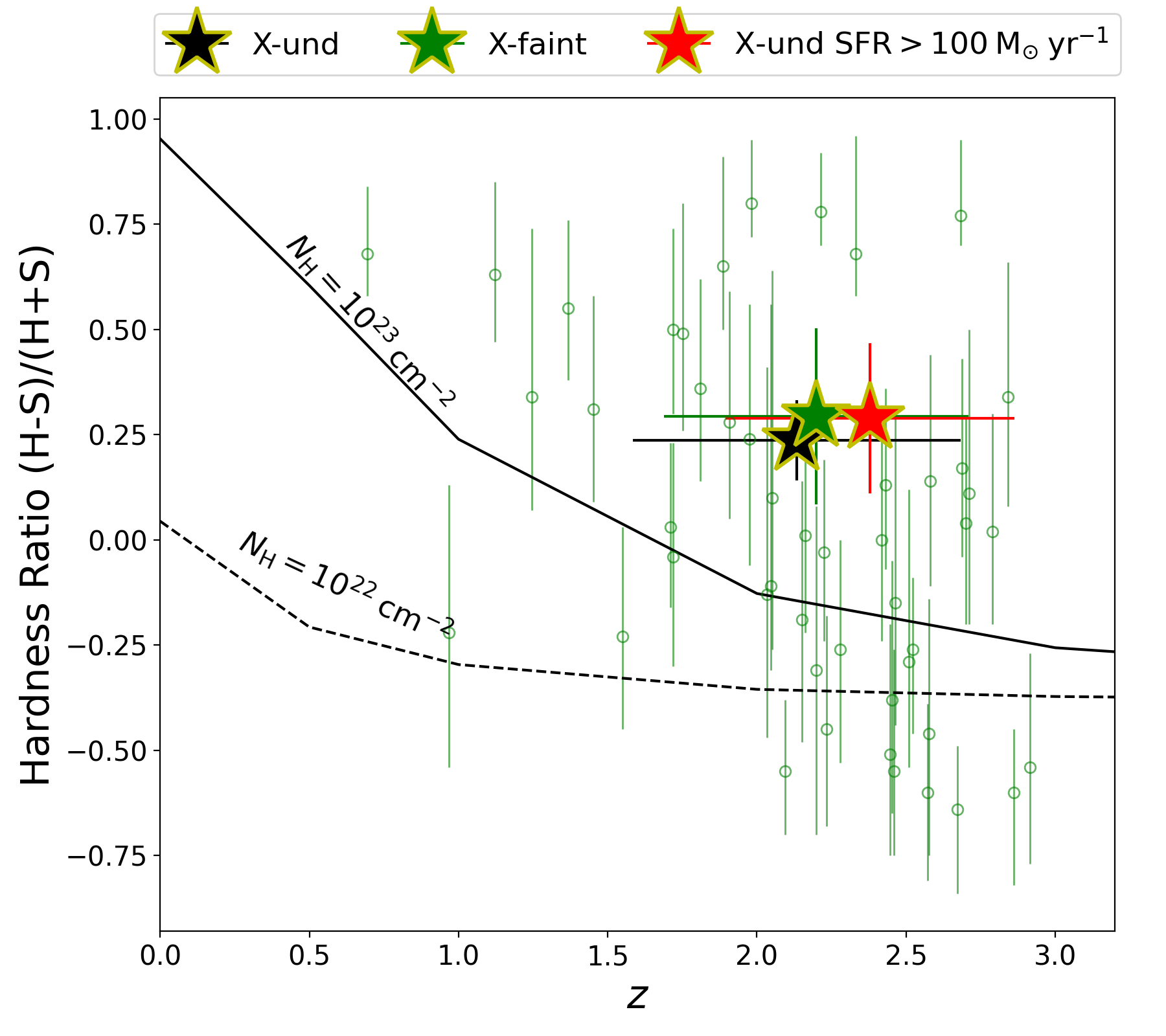}
\caption{Obscuration properties of the IR quasars. {\it Left}: line-of-sight column density ($N_{\rm H}$) against redshift for IR quasars with X-ray spectral fitting constraints. \textit{Right}: Hardness ratio (HR) against redshift for X-ray faint and undetected quasars. Green dots with error bars are the HR of 48 X-ray faint IR quasars provided by \citet{2016bMarchesi}. The black, green, and red stars represent the average HR of the X-ray undetected, X-ray faint, and X-ray undetected quasars with $SFR>100\rm \: M_{\odot} \: yr^{-1}$, respectively, computed after stacking the three samples. The X-ray stacking results are reported in Table\,\ref{t:HR}. The dotted black and continuous lines represent the change of the HR with redshift for $N_{\rm H}=10^{22}\rm \: cm^{-2}$ and $N_{\rm H}=10^{23}\rm \: cm^{-2}$, respectively. The curves were calculated using PIMMS v4.11a, assuming $\Gamma=1.8$, and with observational parameters from {\it Chandra}-Cycle 14.  }
\label{fig:obs_z}
\end{figure*}

Figure\,\ref{fig:obs_z} shows the X-ray obscuration properties of the IR quasars. The left panel depicts the line of sight column density against redshift for the X-ray bright objects (see Section\,\ref{sec:Xraydata} for details on X-ray spectral fitting). Overall, we find that 99 quasars are obscured by $N_{\rm H}\geq 10^{22} \rm \: cm^{-2}$, where 56 are also heavily obscured by $N_{\rm H}> 10^{23} \rm \: cm^{-2}$, of which 6 reside in the CT regime. The remaining 166 X-ray bright quasars are unobscured ($N_{\rm H} < 10^{22} \rm \: cm^{-2}$).

We can estimate the average level of obscuration of the X-ray faint and undetected sources through X-ray stacking. We stack both samples using \textsc{CSTACK} v4.32 \footnote{CSTACK (http://cstack.ucsd.edu/ or http://lambic.astrosen.unam.mx/cstack/) developed by Takamitsu Miyaji}. We find that the average count rate in the 2-8 keV range is  $(1.44\pm 0.21)\times 10^{-5}\rm \: counts \, s^{-1}$ for the X-ray undetected sources, and is $(4.63\pm 1.32)\times 10^{-5}\rm \: counts \, s^{-1}$ for the X-ray faint sources. Figure\,\ref{fig:Xstack} in the Appendix shows the stacked images for X-ray undetected and faint sources, respectively. We detect a strong signal in the hard band (2-8 keV) and a fainter but significant signal in the soft band (0.5-2 keV), where the absorption would have the biggest impact, suggesting that the individual non-detections are due to heavily obscured column densities and not because they are intrinsically faint. 

To estimate the obscuration of the X-ray faint and undetected sources, we use the hardness ratio (HR), which is defined as $\rm HR = \frac{H-S}{H+S}$, where H and S are the net counts in the hard (2-8 keV) and soft bands (0.5-2 keV), respectively \citep{2006Park}. The right panel of Figure\,\ref{fig:obs_z} shows the average HR against redshift of X-ray faint and undetected IR quasars, calculated after stacking the two samples. This panel also shows the HR for the 48 X-ray faint quasars detected in the hard and soft bands from \citet{2016Civano}, together with the expected change in HR with redshift for a fixed column density and intrinsic X-ray spectral slope ($\Gamma=1.8$). Dotted and continuous black lines are the curves for $N_{\rm H}=10^{22}\rm \: cm^{-2}$ and $N_{\rm H}=10^{23}\rm \: cm^{-2}$, respectively, which were calculated using PIMMS v4.11a\footnote{See https://heasarc.gsfc.nasa.gov/cgi-bin/Tools/w3pimms/w3pimms.pl}, for Chandra-Cycle 14 \footnote{We choose to use the Chandra-Cycle 14 response files as that corresponds to the period when the majority of the Chandra observations for the \textit{Chandra} COSMOS Legacy survey were taken.}. The majority of the X-ray faint sources have HR measurements consistent with them being heavily obscured. Additionally, the HR of the stacked X-ray faint and undetected sources indicates that, on average, both samples are heavily obscured by $N_{\rm H} \geq 10^{23}\rm \: cm^{-2}$, with potentially a substantial fraction of CT AGN. Table\,\ref{t:HR} lists the count rates of the stacked X-ray faint and undetected samples.

\begin{table*}
\centering
\scalebox{0.8}{
\begin{tabular}{l c c c c c c c c l} 
 \hline
 \hline
\noalign{\smallskip}
& N & S & H & HR & $z$ & ${\log} L_{6_{\mu \rm m}}$ &  $F_{2-10 \rm \: keV}$ &  ${\log } L_{2-10 \rm \: keV}$ \\ 
&  & ($\rm counts \: s^{-1}$) & ($\rm counts \: s^{-1}$) & &  & ($\rm erg\: s^{-1}$) &  ($\rm erg\: cm^{-2}\: s^{-1}$) &  ($\rm erg\: s^{-1}$) \\ 
 (1) & (2)  & (3) & $(4)$ & (5) & (6) & (7) &(8) &(9) \\ 
\hline
\noalign{\smallskip}
X-ray undetected  & 185 & $(8.88 \pm 1.25) \times 10^{-6}$ & $(1.44 \pm 0.21) \times 10^{-5} $  & $0.24 \pm 0.1$ & $2.13 \pm 0.50$  & $44.93\pm 0.34$ & $(9.2\pm 1.61)\times 10^{-17}$ & $42.41\pm 0.18$ \\ 
\noalign{\smallskip}

\noalign{\smallskip}
X-ray faint  & 71 & $(2.53 \pm 0.97) \times 10^{-5}$ & $(4.63 \pm 1.32) \times 10^{-5} $  & $0.28 \pm 0.21$ & $2.14 \pm 0.51$ & $45.01\pm 0.31$ & $(2.62\pm 1.11)\times 10^{-16}$ & $42.87\pm 1.00$\\ 
\noalign{\smallskip}

X-ray undetected $\&$ $\rm SFR > 100 \: M_{\odot} \: yr^{-1} $  & 62 & $(8.11 \pm 2.3) \times 10^{-6}$ & $(1.47 \pm 0.40) \times 10^{-5} $  & $0.29 \pm 0.18$ & $2.30 \pm 0.48$ & $45.02\pm 0.33$ & $(8.41\pm 2.71)\times 10^{-17}$ & $42.45\pm 0.32$\\ 

\noalign{\smallskip}
\hline
\hline

\end{tabular}}
\caption{X-ray stacking results for the X-ray undetected, X-ray faint ($<30$ net counts), and X-ray undetected sources with $SFR>100\rm \: M_{\odot} \: yr^{-1}$. (1) The sample; (2) number of stacked sources; (3) net soft band count rates (S, 0.5-2 keV) (4) net hard band count rates (H, 2-8 keV); (5) the hardness ratio ($\rm HR = (H-S)/(H+S)$); (6) the mean redshift of the subsamples; (7) mean and standard deviation of AGN rest-frame $6 \: \mu \rm m$ logarithmic luminosity: (8) mean 2-10 keV flux estimated from the 0.5-2 keV stacked count rates; (9) mean $\log L_{\rm 2-10 \: keV}$ calculated from $F_{\rm 2-10 \: keV}$. }

\label{t:HR}
\end{table*}

\begin{figure}
\hspace{-0.5cm}
\includegraphics[scale=0.39]{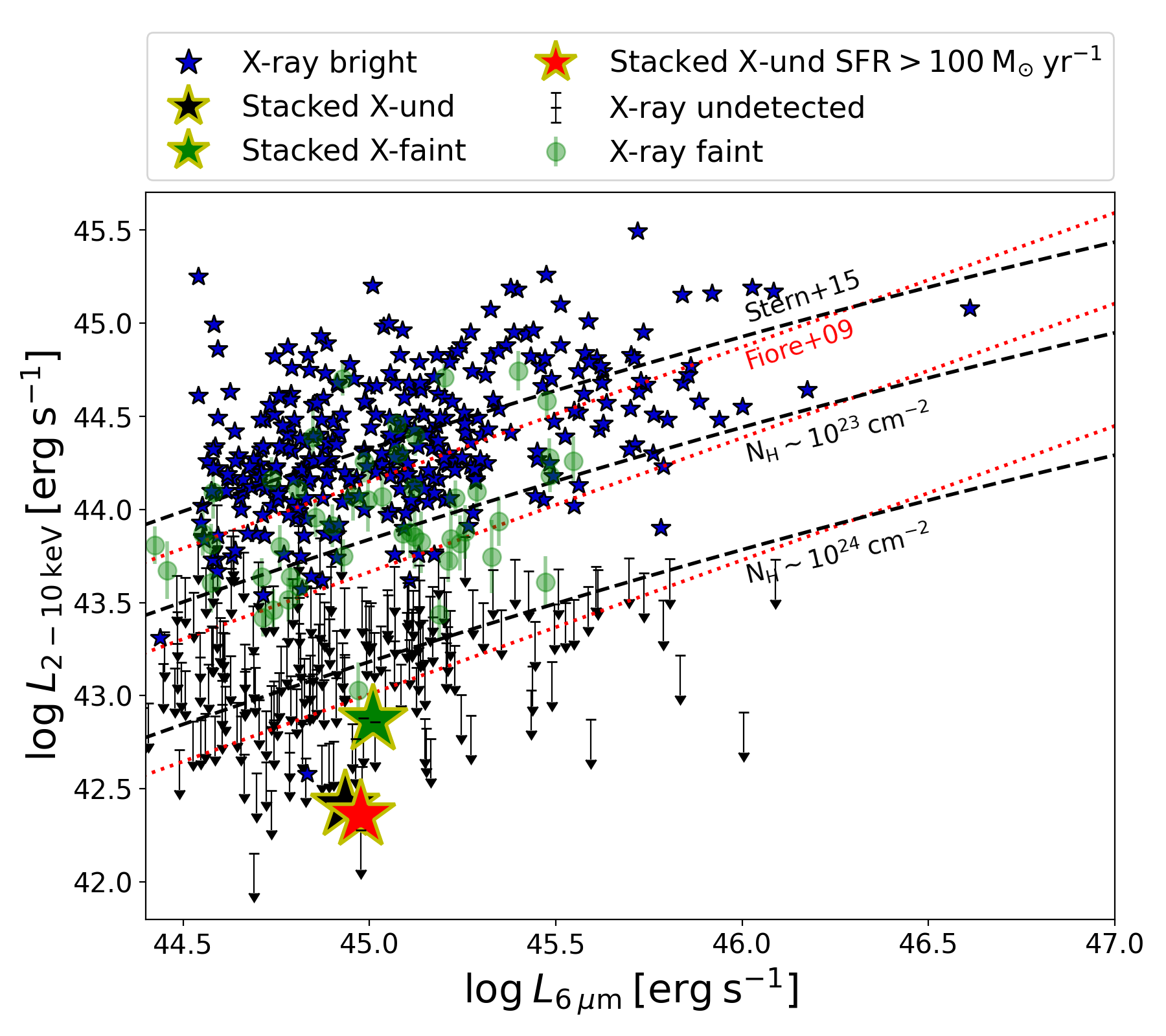}

\caption{Intrinsic rest-frame 2--10 keV luminosity ($L_{\rm 2-10\rm keV}$) against the AGN rest-frame $\rm 6\: \mu m$ luminosity ($L_{\rm 6\: \mu m}$) for the IR quasar sample, computed from the SED fits. Blue stars correspond to the intrinsic $L_{\rm 2-10\rm keV}$ for IR quasars with X-ray spectral fitting results, green circles with error bars correspond to the observed $L_{\rm 2-10\rm keV}$ of the X-ray faint quasars, which are lower limits of the intrinsic $L_{\rm 2-10\rm keV}$, and black upper limits are 3$\sigma$ luminosity upper limits for the X-ray undetected quasars. The upper dashed black and dotted red lines show the intrinsic $L_{\rm 2-10\rm keV}$ -- $L_{\rm 6\: \mu m}$ relation found by \citet{2015Stern} and \citet{2009Fiore}, respectively. The lower dashed black and dotted red lines show the same relations but including the impact of $N_{\rm H} \sim 10^{23} \rm \: cm^{-2}$ and $N_{\rm H} \sim 10^{24} \rm \: cm^{-2}$ on the observed $L_{\rm 2-10\rm keV}$; these are calculated assuming an absorbed power-law with photon index $\Gamma=1.8$ using PIMMS v4.11a. The green, black, and red stars represent the average position of X-ray faint, X-ray undetected, and X-ray undetected sources with star-formation rates (SFR) $\rm SFR>100\: M_{\odot} \: yr^{-1}$ (see Sections\,\ref{sec:res:SFR} and Appendix\,\ref{sec:res:obsXraynoXray} for details), respectively, calculated after stacking the three samples. }
\label{fig:Lx_L6um}
\end{figure}

In addition to the direct $N_{\rm H}$ constraints from the X-ray spectral fitting, and the $N_{\rm H}$ estimated from the HR, we can further estimate the broad level of obscuration of the X-ray faint and undetected quasars using the known relation between the intrinsic $L_{2-10 \rm keV}$ and the AGN rest-frame $6 \: \mu \rm m$ luminosity ($L_{6\rm \mu m}$) \citep[e.g.,][]{2004Lutz,2009Fiore,2015Stern}. Figure\,\ref{fig:Lx_L6um} depicts the intrinsic $L_{2-10 \rm keV}$ against $L_{6\rm \mu m}$ of the IR quasar sample, together with the intrinsic $L_{2-10 \rm keV}$--$L_{6\rm \mu m}$ relations of \citet{2015Stern} and \citet{2009Fiore}. For the 185 IR quasars that are X-ray undetected, we calculate 3-$\sigma$ upper limits for the observed $2-10\rm \: keV$ flux as described in Section\,\ref{sec:Xuplims}. For the X-ray faint quasars, we use the 2-10 keV observed flux reported in \citet{2016Civano}. The luminosities and upper limits are calculated as $L_{\rm 2-10 keV} = 4 \pi d_{\rm L}^{2} f_{\rm 2-10keV}(1+{\rm z})^{\rm \Gamma -2} \rm \: erg \: s^{-1}$, assuming $\Gamma=1.8$ \citep{2003Alexander}. The upper limits for the observed $L_{2-10 \rm keV}$ of the undetected quasars are shown in black and the majority of them lie below the $L_{2-10 \rm keV}$--$L_{6\rm \mu m}$ relation, in the region of the plane affected by a CT obscuration of $N_{\rm H} \sim 10^{24} \rm \: cm^{-2}$. 

In order to locate the average position of X-ray faint and undetected quasars in the $L_{2-10 \rm \: keV}$--$L_{6\rm \mu m}$ plane, we use the count rates calculated from the X-ray stacking analysis to estimate the average observed  $L_{2-10 \rm \: keV}$ of both samples. We convert the count rates to flux using PIMMS, assuming $\Gamma=1.8$ and considering the average Galactic absorption in the direction of the COSMOS field ($N_{\rm H} = 2.5 \times 10^{20} \rm \: cm^{-2}$, \citealp[]{2005Kalberla}). We estimate a $2-10\rm \: keV$ flux of $F_{2-10\rm \: keV,und}=9.2\times 10^{-17} \rm \: erg\: cm^{-2}\: s^{-1}$ for the X-ray undetected quasars and $F_{2-10\rm \: keV,faint}=2.62\times 10^{-16} \rm \: erg\: cm^{-2}\: s^{-1}$ for the X-ray faint quasars. Then, we convert the flux to luminosity considering the average redshift of the X-ray faint and undetected sources ($z_{\rm faint}=2.14$ and $z_{\rm und}=2.13$). The green and black stars on Figure\,\ref{fig:Lx_L6um} mark the average location of X-ray faint and undetected quasars, respectively, where the x-axis positions are the average values of $L_{6\rm \mu m}$ (${\rm log \:}L_{6\rm \mu m,faint}/{\rm \: erg \: s^{-1} } = 45.01 $ and ${\rm log \:} L_{6\rm \mu m, und}/{\rm \: erg \: s^{-1} }=  44.93$). The X-ray stacked count rates suggest that X-ray faint and undetected quasars are heavily obscured by $N_{\rm H} \gtrsim 10^{24} \rm \: cm^{-2}$.

With all the evidence presented above (i.e., X-ray stacking, HR, and $L_{2-10 \rm \: keV}$--$L_{6\rm \mu m}$ analysis), we are confident that the samples of X-ray faint and undetected quasars are a good representation of heavily obscured AGNs and probably comprise the most obscured subset of the IR quasar population in the COSMOS field, with CT column densities in many cases. Therefore, we group all the X-ray faint and undetected quasars with the X-ray bright quasars with $N_{\rm H} \geq 10^{22} \rm \: cm^{-2}$ to create an obscured IR quasar sample, which is composed of 355 sources ($\approx$~61\% of the IR-quasar sample). All the quasars with $N_{\rm H} < 10^{22} \rm \: cm^{-2}$ are grouped into an unobscured IR quasar sample, composed of 223 sources ($\approx$~39\% of the IR-quasar sample).

\subsection{Star formation rates} \label{sec:res:SFR}

We analyze the star formation properties of the IR quasar sample using the results of our SED fitting analysis. We directly measure the $L_{\rm SF, IR}$ for 183 quasars, while for the remaining 395 objects, we provide an upper limit (see Section\,\ref{sec:res:SFR} for details). The large fraction (68\%) of upper limits is due to two main reasons: (1) 35\% of the sample does not have {\it Herschel} detections at 2$\sigma$, and (2) our sample contains powerful quasars, for which the AGN component dominates the SED in most cases. 

We start by comparing the SFR of the obscured and unobscured quasars. We convert the  $L_{\rm SF, IR}$ that we obtain from our SED fitting to SFR using the \citet{1998Kennicutt} relation. To remove a possible dependency of the SFR on the AGN luminosity and redshift, we match the unobscured and obscured samples in $L_{6\mu \rm m}$ and redshift. In each matching iteration, for each unobscured quasar, we randomly choose and assign an obscured quasar with a luminosity and redshift that lie within $\Delta {\rm log \:}L_{6\mu \rm m}=0.2 \: \rm dex$ and $\Delta z=0.05\times (1+z)$, respectively. The resulting matched samples consist of 250 unobscured and 250 obscured IR quasars, with similar redshifts and AGN luminosity. 

\begin{figure}
\centering
\includegraphics[scale=0.42]{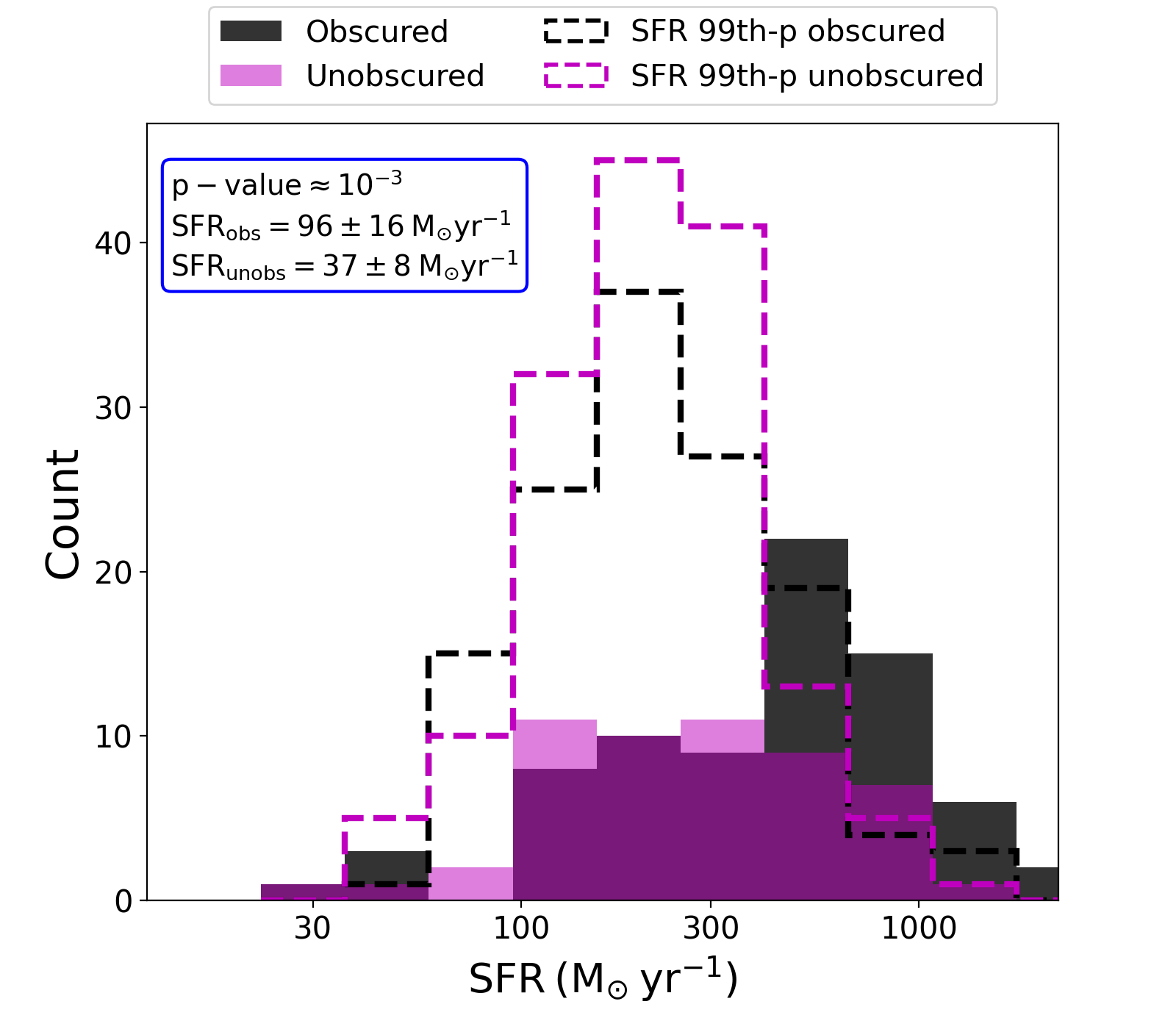}

\caption{Distribution of the SFR calculated from the $8-1000\: \mu \rm m$  SF luminosity of unobscured (magenta) and obscured (black) sources. The filled and empty histograms are the median values of the well constrained SFRs, and the dashed empty histograms are the 99th percentile of the posterior distribution (i.e., $2.3\sigma$ upper-limits). The plot was made after matching the unobscured and obscured samples in redshift and AGN $6 \: \mu \rm m$ luminosity to remove any luminosity biases. The values quoted on the figure correspond to the mean SFRs calculated with \textsc{PosteriorStacker}, including the upper-limit values (see Sections\,\ref{sec:uplims} and\,\ref{sec:res:SFR} for details) of the matched samples. The histograms are not normalized. }
\label{fig:LSF_dist}
\end{figure}

Figure\,\ref{fig:LSF_dist} shows the distribution of SFR for the unobscured and obscured IR quasars. From the plot, it is clear that there are more obscured quasars with well constrained SFR than unobscured quasars, and that the obscured sample has a larger population of high SFR ($\gtrsim 300 \rm \: M_{\odot} \: yr^{-1}$) quasars than the unobscured sample. As the luminosity-redshift matching is performed with randomization, we repeat the process 100 times and compute the number of $L_{\rm SF, IR}$ upper limits for unobscured and obscured sources. We find that, on average, $\approx 63\%$ of the obscured sources have a $L_{\rm SF, IR}$ upper limit, and that this number increases to $\approx 75\%$ for unobscured sources. We also investigated the SFR distributions for the systems where the SF component is directly measured (i.e., a well constrained $L_{\rm SF, IR}$) by applying the Kolmogorov-Smirnov (KS) test for the unobscured and obscured samples. We find an average $\rm p-value \approx 0.001$, implying that the two distributions are different at the $>99\%$ confidence level. We also calculate the average $L_{\rm SF, IR}$ and SFR of the unobscured and obscured matched samples. In order to account for the upper limits, we use \textsc{PosteriorStacker}. We find that for unobscured quasars, the mean $\log L_{\rm SF, IR}$ is $\mu_{unobs}=44.87\pm 0.1 \: \rm erg \: s^{-1} $ and the standard deviation is $\sigma_{unobs}=0.80\pm 0.08\:\rm erg \: s^{-1}$, while for obscured quasars, $\mu_{obs}=45.34\pm 0.1\rm \:erg \: s^{-1}$ and $ \sigma_{obs}=0.7\pm 0.06\: \rm erg \: s^{-1}\rm \: erg \: s^{-1}$. The mean SFRs are $\rm SFR_{obs }=98\pm 16 \rm \: M_{\odot} \, yr^{-1}$ and $\rm SFR_{unobs }=33\pm 8 \rm \: M_{\odot} \, yr^{-1}$, indicating that the obscured quasars have a $\rm SFR \approx 3 \pm 1$ times higher than unobscured quasars. 

To complement the SED fitting results, we also calculate the number of {\it Herschel} detections and their associated binomial errors in both samples for 100 matching realizations. We found that $73\pm 0.002\%$ ($53\pm 0.002 \%$) of the obscured quasars have at least one (two) $2\sigma$ detection in one (two) of the {\it Herschel} bands, while this number decreases to $49\pm  0.002\%$ ($35\pm 0.003\%$) for unobscured quasars. The higher {\it Herschel} detection fraction in obscured quasars is further confirmation of them hosting larger SFRs than unobscured quasars.

The increased SFRs of our obscured quasar sample could be due to strongly star-forming galaxies masquerading as obscured quasars, particularly within the X-ray undetected subset; i.e.,\ our SED fitting may have wrongly identified an IR-AGN component. Overall, 62 of the 170 quasars with $SFR>100\rm \: M_{\odot} \: yr^{-1}$ are X-ray undetected. We stacked the X-ray data of these X-ray undetected strongly star-forming systems, which reveals significant hard X-ray emission with a HR and basic properties consistent with the overall X-ray undetected and X-ray faint samples; see Appendix\,\ref{sec:res:obsXraynoXray}, Table\,\ref{t:HR}, Figure\,\ref{fig:obs_z}, and Figure~\ref{fig:Lx_L6um}. These analyses show that the X-ray undetected high SFR systems are obscured quasars with properties consistent with the X-ray bright, faint, and undetected quasars.

Table\,\ref{t:results} shows a summary of the mean $L_{\rm SF, IR}$, SFR, and {\it Herschel} detections of the obscured and unobscured quasar samples, quantifying the evidence for a significant SF enhancement in the obscured quasars compared to the unobscured quasars.

\begin{table*}
\hspace{-0.45cm}
\scalebox{0.72}{
\begin{tabular}{p{12mm} c c c p{19mm} c c c c c c c p{19mm}} 
 \hline
 \hline

\noalign{\smallskip}
 Sample & & N & ${\log} L_{\rm 6\mu m}$ & ${\log} L_{\rm SF, IR}$ &SFR &  $\geq 1$ {\it Herschel} & $\geq 2$ {\it Herschel} & $\mathcal{R}$ & $\mathcal{R}_{\rm SF}$ & $\mathcal{R}_{\rm AGN}$ & $f_{\rm VLA}$ &  ${\log} M_{\star}$ \\  
\noalign{\smallskip}
 &  & & ($\rm erg \: s^{-1}$) & ($\rm erg \: s^{-1}$) & $(\rm M_{\odot}\: yr^{-1})$ &  & & & & & &  ($M_{\odot}$) \\ 
 (1) & & (2)  & (3) & $(4)$ & (5) & (6) & (7) &(8) &(9) & (10)& (11) & (12)  \\ 
\noalign{\smallskip}
\hline
\noalign{\smallskip}
IR quasars & Unobscured & 205 & $45.03 \pm 0.05$ & $\mu = 44.87 \pm 0.1$ & $\mu= 33\pm8$  & $0.53\pm 0.002$ & $0.35\pm 0.003$  & $-4.85 \pm 0.01$ & $ -5.1\pm 0.01$ & $-4.65 \pm 0.02$ & $0.68\pm 0.002$ & $\mu = 10.86 \pm 0.02$  \\ 
 & & & & $\sigma=0.8\pm 0.08$  &  & & & & & & & $\sigma=0.21\pm 0.02$ \\ 
\noalign{\smallskip}

\noalign{\smallskip}
 & Obscured & 205 & $45.05\pm 0.05$ & $\mu = 45.34 \pm 0.1$  & $\mu= 98\pm16$ & $0.73 \pm 0.002$ &$ 0.49\pm 0.002$ & $-4.66 \pm 0.02$ & $-4.86\pm 0.02$ & $-4.56\pm 0.03$ & $0.77\pm 0.002$  & $\mu = 10.71 \pm 0.02$ \\ 
 & & & & $\sigma=0.7\pm 0.06$  &   & & & & & &  & $\sigma=0.30\pm 0.02$  \\ 
\noalign{\smallskip}
\hline 
\noalign{\smallskip}
Obscured \newline IR quasars & X-ray bright  & 85 & $45.01 \pm 0.05$ & $\mu = 45.12 \pm 0.12$ \newline $\sigma=0.74\pm 0.10$ & $ \mu = 69\pm19$  & $0.72\pm 0.003$ & $0.48\pm 0.004$  & $-4.71 \pm 0.01$ & $-4.90 \pm 0.02 $ & $ -4.61\pm 0.01$ & $0.79\pm 0.03$ &$\mu = 10.8 \pm 0.04 $ \newline $\sigma=0.32\pm 0.03$  \\ 
 \noalign{\smallskip}
& X-ray faint/und  & 85 & $45.03 \pm 0.05$ & $\mu = 45.29 \pm 0.11$ \newline $\sigma=0.70\pm 0.10$ & $ \mu = 88 \pm 22 $  & $0.75\pm 0.003$ &$0.49\pm 0.004$  & $-4.70 \pm 0.01$ & $ -4.87\pm 0.03 $ & $ -4.56 \pm 0.05$ & $0.78\pm 0.003$ & $\mu =10.7 \pm 0.03$ \newline $\sigma=0.24\pm 0.03$   \\ 
\noalign{\smallskip}
\hline
\hline

\end{tabular}}
\caption{Mean AGN and host galaxy properties of our sample {bf matched in $z$ and $L_{\rm 6 \mu \rm m}$}. The first two rows show the properties of the obscured and unobscured IR quasar samplesand the last two rows show the properties of the obscured X-ray bright and X-ray faint/undetected quasars. We report the mean values for each property obtained after matching the samples 100 times. Columns} (1) the sample; (2) the number of quasars in each matched sample; (3) mean AGN rest-frame $6\rm \mu m$
luminosity; (4) mean ($\mu$) and standard deviation ($\sigma$) of the star-formation luminosity distribution of the sample calculated with \textsc{PosteriorStacker}; (5) mean SFR estimated from the mean star-formation luminosity; (6) fraction of the sample with at least one $2\sigma$ {\it Herschel} detection; (7) fraction of the sample with at least two $2\sigma$ {\it Herschel} detections; (8) mean and standard deviation of the radio-loudness distribution of the sample; (9) same as (8) for sources where the radio emission is dominated by star-formation processes adopting $q_{\rm TIR}=2.64$ \citep[][]{2003Bell}; (10) same as (9) for sources where the radio emission is dominated by AGN processes; (11) VLA radio detection fraction; (12) mean ($\mu$) and standard deviation ($\sigma$) of the stellar mass distribution of the sample calculated with \textsc{PosteriorStacker}.  
\label{t:results}
\end{table*}

\subsection{Stellar Masses} \label{sec:res:Mstar}

\begin{figure}
\centering
\includegraphics[scale=0.42]{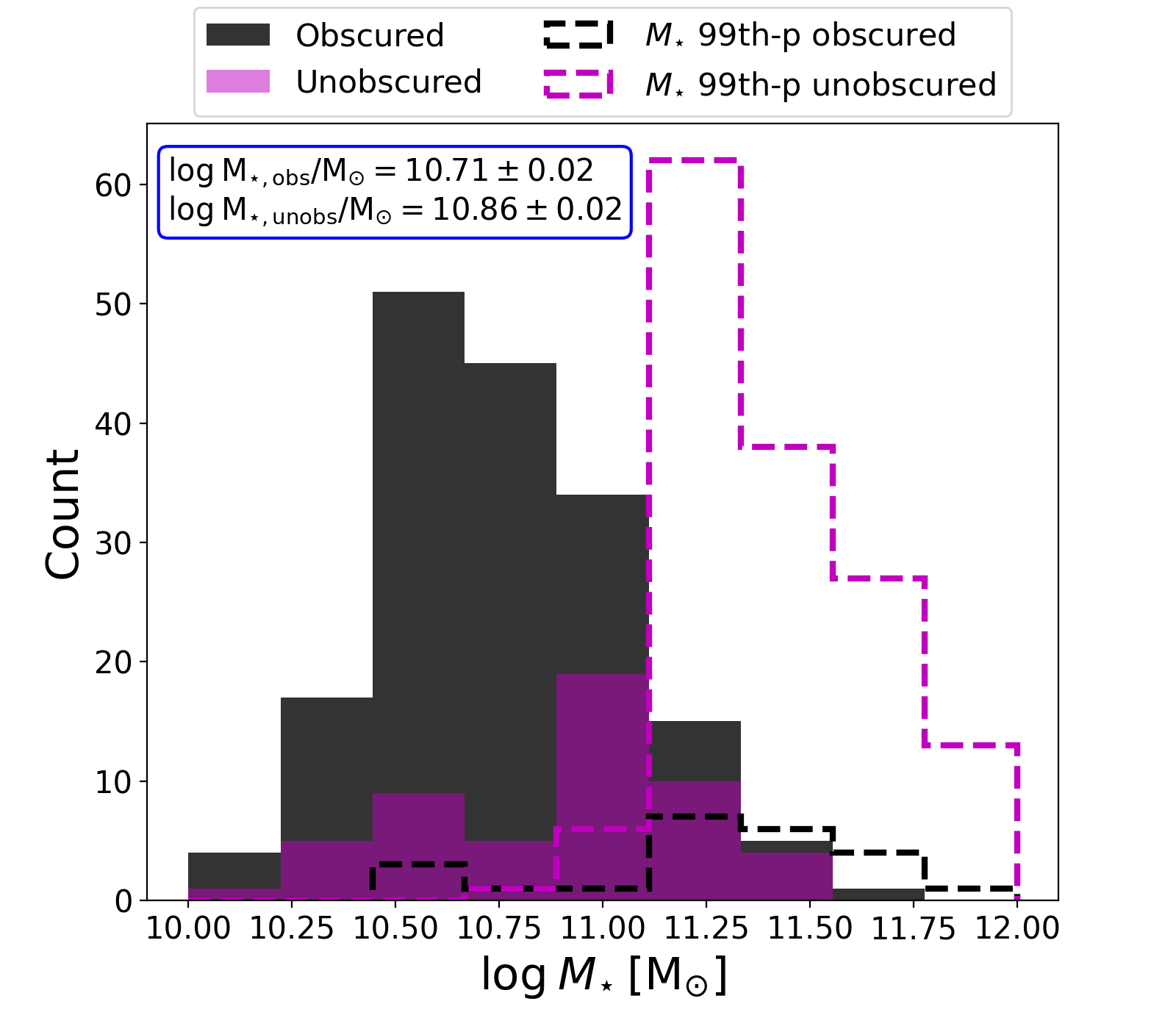}
\caption{Distribution of stellar masses for obscured (black) and unobscured (magenta) IR quasars. The filled histograms are the median values of the well constrained stellar masses, and the dashed empty histograms are the 99th percentile of the posterior distribution. The plot was made after matching the unobscured and obscured samples in redshift and AGN $6\mu \rm m$ luminosity to remove any luminosity biases. The values quoted on the figure correspond to the mean stellar masses calculated with \textsc{PosteriorStacker} including the upper-limit values (see Sections\,\ref{sec:uplims} and\,\ref{sec:res:SFR} for details) of the matched samples. The histograms are not normalized. } 

\label{fig:Mstar}
\end{figure}

Our SED fitting also provides constraints on the stellar masses of the IR quasars. For the obscured sample, we can constrain the stellar masses for the majority ($87\%$) of the sources. For the unobscured sample, we can only provide well-constrained masses for $26\%$ of the quasars, since it is more difficult to reliably constrain the stellar population emission due to the significant wavelength overlap with the accretion disk emission. 

Figure\,\ref{fig:Mstar} shows the stellar mass distributions for the obscured and unobscured IR quasars after matching the samples in redshift and AGN $6\rm \: \mu m$ luminosity. We calculate the mean stellar masses of obscured and unobscured quasars using \textsc{PosteriorStacker} (i.e., including unconstrained values) and find that, for obscured quasars, the mean and standard deviation are $\log M_{\rm \star, obs}/{\rm M_{\odot}} = 10.71\pm 0.02 $ and $\sigma_{\rm M_{\star}, obs} = 0.30 \pm 0.02 \:{\rm M_{\odot}}$, respectively, and that for unobscured quasars $\log M_{\rm \star, unobs}/{\rm M_{\odot}} = 10.86 \pm 0.02 $ and $\sigma_{\rm M_{\star}, unobs} = 0.21 \pm 0.02 \:{\rm M_{\odot}}$. The difference between the mean stellar masses of both samples is around $0.15 \rm \: dex$, indicating that unobscured quasars have slightly more massive host galaxies. However, we should not overinterpret this small difference since the mean stellar mass of unobscured quasars is dominated by the prior adopted in the SED fitting. We note that the dispersion of the stellar mass distribution of obscured quasars is higher than for the unobscured quasars. This is probably because we do not apply a prior to fit the stellar masses of the obscured quasars, while the fit is significantly influenced by the prior for the majority of the unobscured quasars. 

As a check, we compare the stellar masses of our IR quasars calculated in this work with the values reported in the COSMOS 2015 catalog \citep{2016Laigle}. We find broad agreement in the stellar masses with COSMOS 2015 for the obscured quasars that do not require an AD to model their optical emission (i.e.,\ those where the host galaxy dominates the optical emission). However, we find large inconsistencies for the quasars that need an AD component since in \citet{2016Laigle} the optical emission is solely modelled with a stellar population component without any contribution from an AGN.

The results reported above (summarized in Table\,\ref{t:results}) indicate that unobscured quasars have comparable stellar masses to their obscured counterparts. While the mean stellar masses of the unobscured quasars are influenced by the stellar-mass prior, it is unlikely that the mean stellar mass of the unobscured quasars could be significantly smaller than the obscured quasars. Consequently, these results indicates that the observed difference in the SFRs between obscured and unobscured quasars is not due to the obscured quasars being substantially more massive than the unobscured quasars.

\subsection{Radio properties} \label{sec:res:radio}

\begin{figure*}
\centering
\includegraphics[scale=0.43]{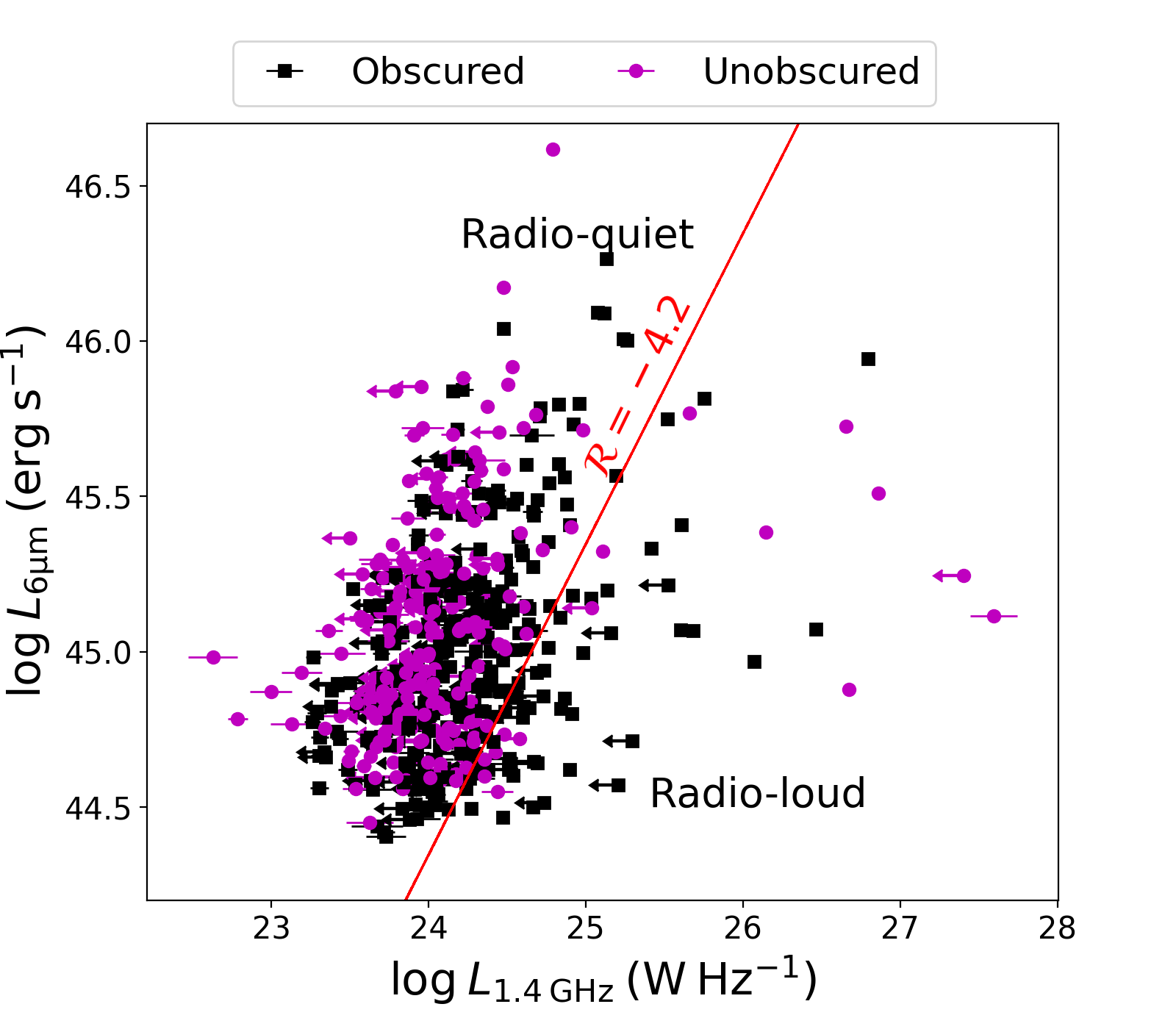}
\includegraphics[scale=0.43]{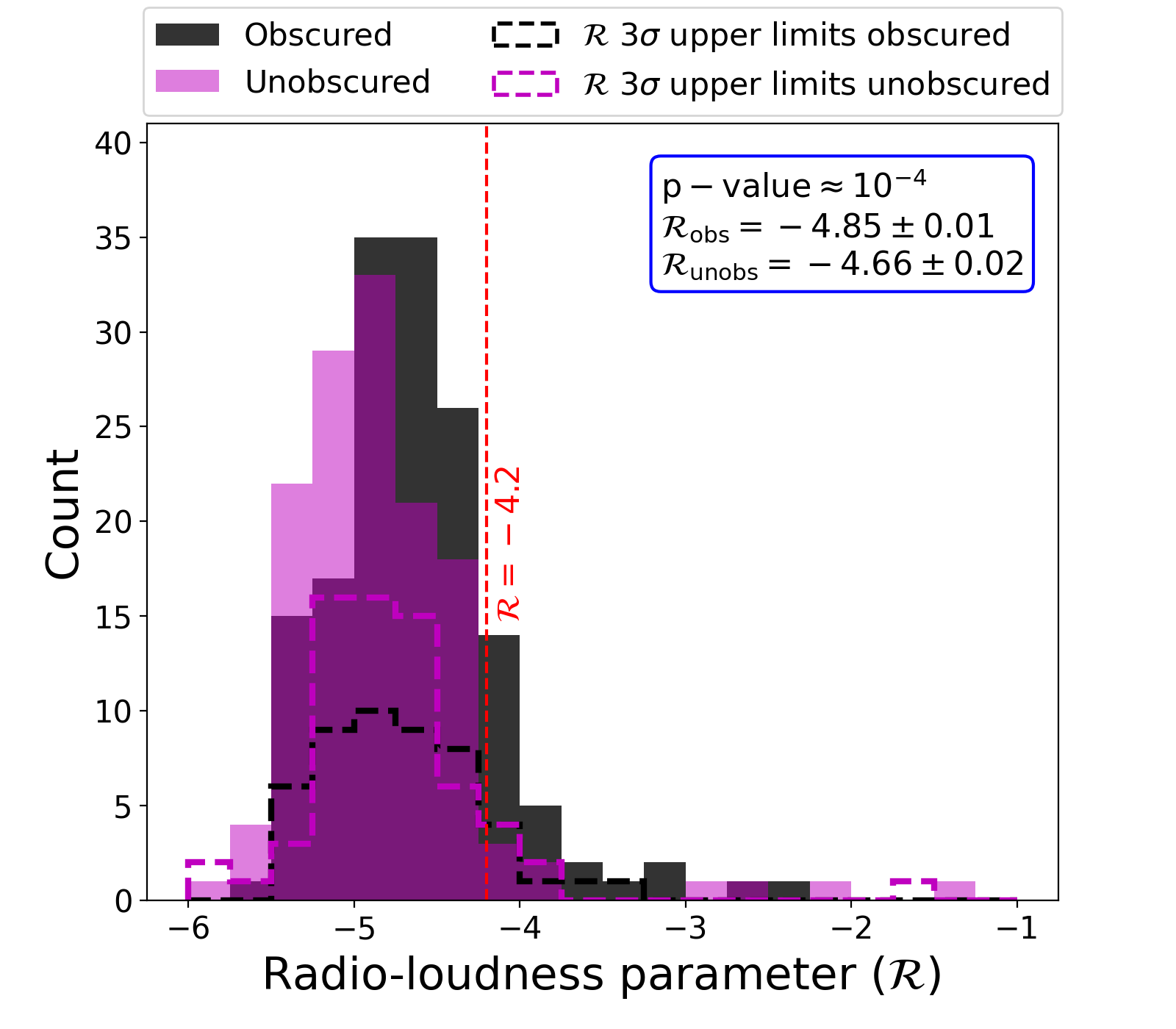}

\caption{ Radio-loudness properties. {\it Left:} AGN rest-frame $6\:\mu \rm m$ luminosity ($L_{\rm 6 \mu  m}$) versus the 1.4 GHz luminosity ($L_{\rm 1.4\: GHz}$). Black squares are obscured IR quasars, and magenta circles are unobscured IR quasars. The red line indicates a radio-loudness of $\mathcal{R}=-4.2$ (see Equation\,\ref{eq:RL}), our adopted radio-loud/radio-quiet threshold. {\it Right:} Distribution of the radio-loudness parameter ($\mathcal{R}$) of unobscured (magenta) and obscured (black) sources. The dotted red line marks $\mathcal{R}=-4.2$, the radio-loud/radio-quiet threshold. The filled histograms are the RL distributions for radio-detected quasars, and the dashed empty histograms are the $3\sigma$ RL upper limits for radio-undetected quasars. The plot was made after matching the unobscured and obscured samples to remove any dependency with the redshift and AGN $6 \: \mu \rm m$ luminosity. The values reported in the figure are calculated after matching the samples 100 times and are the mean values for the $\rm p-value$ of the RL distributions of obscured and unobscured IR quasars, and the mean RL of obscured and unobscured IR quasars. }
\label{fig:L14_L6um}
\end{figure*}

We investigate the radio properties of our IR quasar sample. Overall, 420 out of 578 IR quasars ($73\%$ of the sample) have at least a 2$\sigma$ detection at 3 or 1.4 GHz in the VLA-COSMOS 3GHz survey or \citetalias{2018Jin}. The radio detection fractions ($f_{\rm VLA}$) of unobscured and obscured quasars are $f_{\rm VLA}=70\%$ and $f_{\rm VLA}=75\%$, respectively. We convert the radio fluxes to the individual quasars to rest-frame luminosity by calculating $L_{1.4 \rm \: GHz}=4\pi d_{\rm L}^2 f_{1.4 \rm GHz} 10^{-36}(1+z)^{\alpha-1}$, following \citet{2003Alexander} and assuming a radio spectral index of $\alpha=-0.9$; as described in Section\,\ref{sec:data:radio} we convert 3~GHz fluxes to 1.4~GHz fluxes.

To quantify the strength of the radio emission, we use the radio-loudness (RL) parameter, $\mathcal{R}$, which quantifies the relative strength of the radio emission to the rest-frame $6\mu m$ AGN emission : 

\begin{equation}\label{eq:RL}
\mathcal{R} = \rm \log \frac{1.4 \times 10^{16} L_{1.4 \rm \: GHz} \rm \: [W\: Hz^{-1}]}{L_{\rm 6 \mu m} \rm \: [erg \: s^{-1}]}
\end{equation}

\noindent \citep[e.g.,][]{2019Klindt,2020Fawcett}. We adopt $\mathcal{R}=-4.2$ as the radio-loud/radio-quiet threshold, following the prescription of \citet{2019Klindt}. The left panel of Figure\,\ref{fig:L14_L6um} shows $L_{6\mu \rm m}$ against $ L_{1.4 \rm \: GHz}$ for obscured and unobscured radio detected IR quasars. The red line represents $\mathcal{R}=-4.2$. Most of the quasars reside in the radio-quiet regime. There are some visual differences between the RL distributions of unobscured and obscured quasars: the obscured systems tend to have larger $\mathcal{R}$ values (i.e., higher $ L_{1.4 \rm \: GHz}$ for a similar $L_{6\mu \rm m}$) than the unobscured systems, with an apparent excess of obscured IR quasars above the radio-loud/radio-quiet threshold.

To test whether the enhancement in the radio is genuine, we compare the RL distributions of unobscured and obscured quasars using the matched samples described in Section\,\ref{sec:res:SFR}, in order to remove any possible dependency in redshift and AGN luminosity. The right panel of Figure\,\ref{fig:L14_L6um} shows RL distributions of unobscured (magenta) and obscured (black) sources. 
The distributions appear to be different, with relatively more obscured quasars in the RL interval of [-5, -4]. As in Section\,\ref{sec:res:SFR}, we match the unobscured and obscured samples and perform the KS test 100 times. The mean RL of obscured and unobscured quasars are $\mathcal{R}=-4.66 \pm 0.02$ and $\mathcal{R}=-4.85\pm 0.01$, respectively. Consequently, we can rule out that the two samples are drawn from the same distribution at a $99.99\% $ significance ($\rm p-value\approx 0.001$). We also compute the radio detection fraction for the 100 matched samples, finding that the average radio detection fractions of unobscured and obscured quasars are $f_{\rm VLA}=68\pm 0.02\%$ and $f_{\rm VLA}=77\pm 0.02 \%$, respectively. Both of these results indicate that obscured quasars have stronger radio emission than unobscured quasars. 

\begin{figure*}
\centering
\includegraphics[scale=0.35]{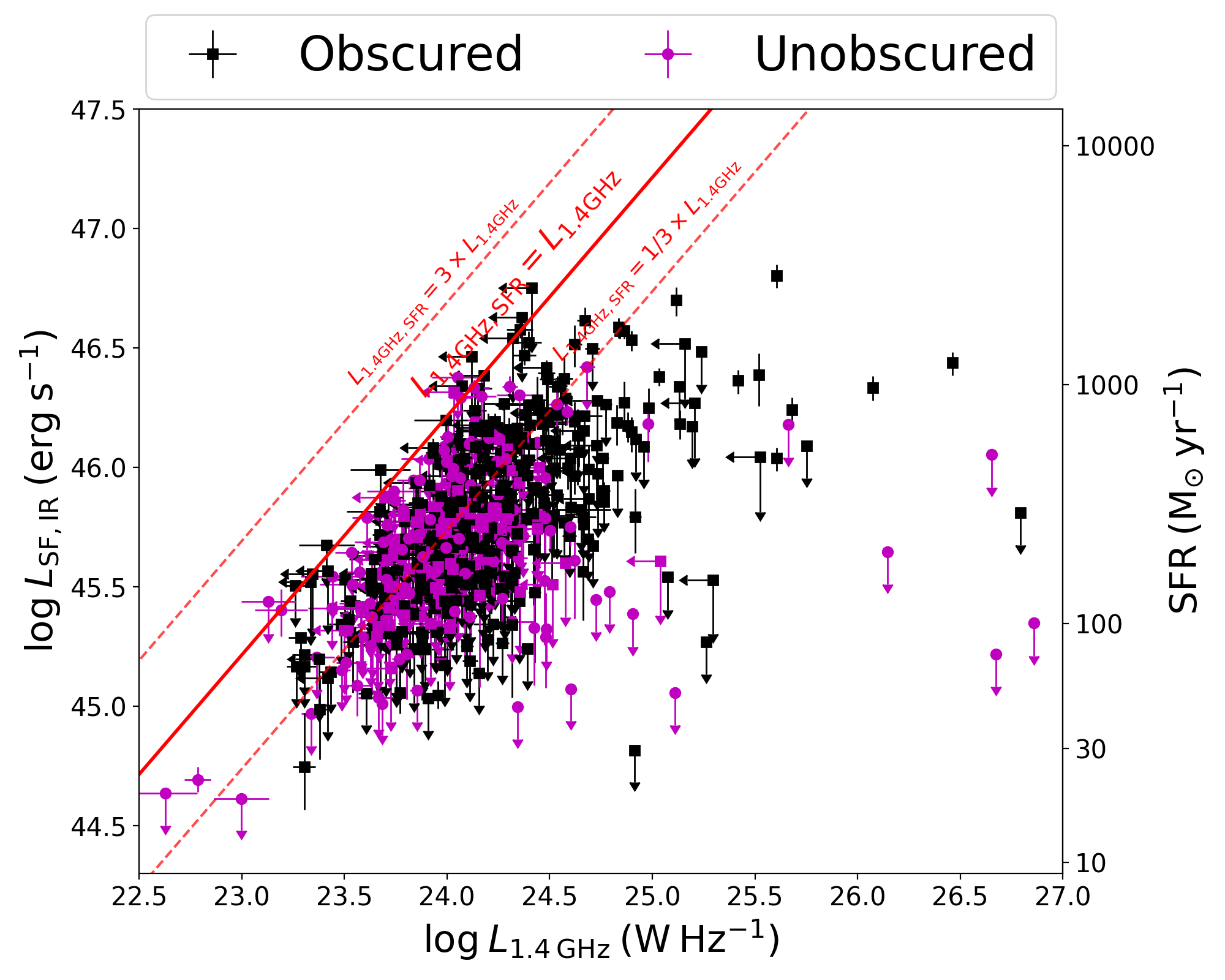}
\includegraphics[scale=0.35]{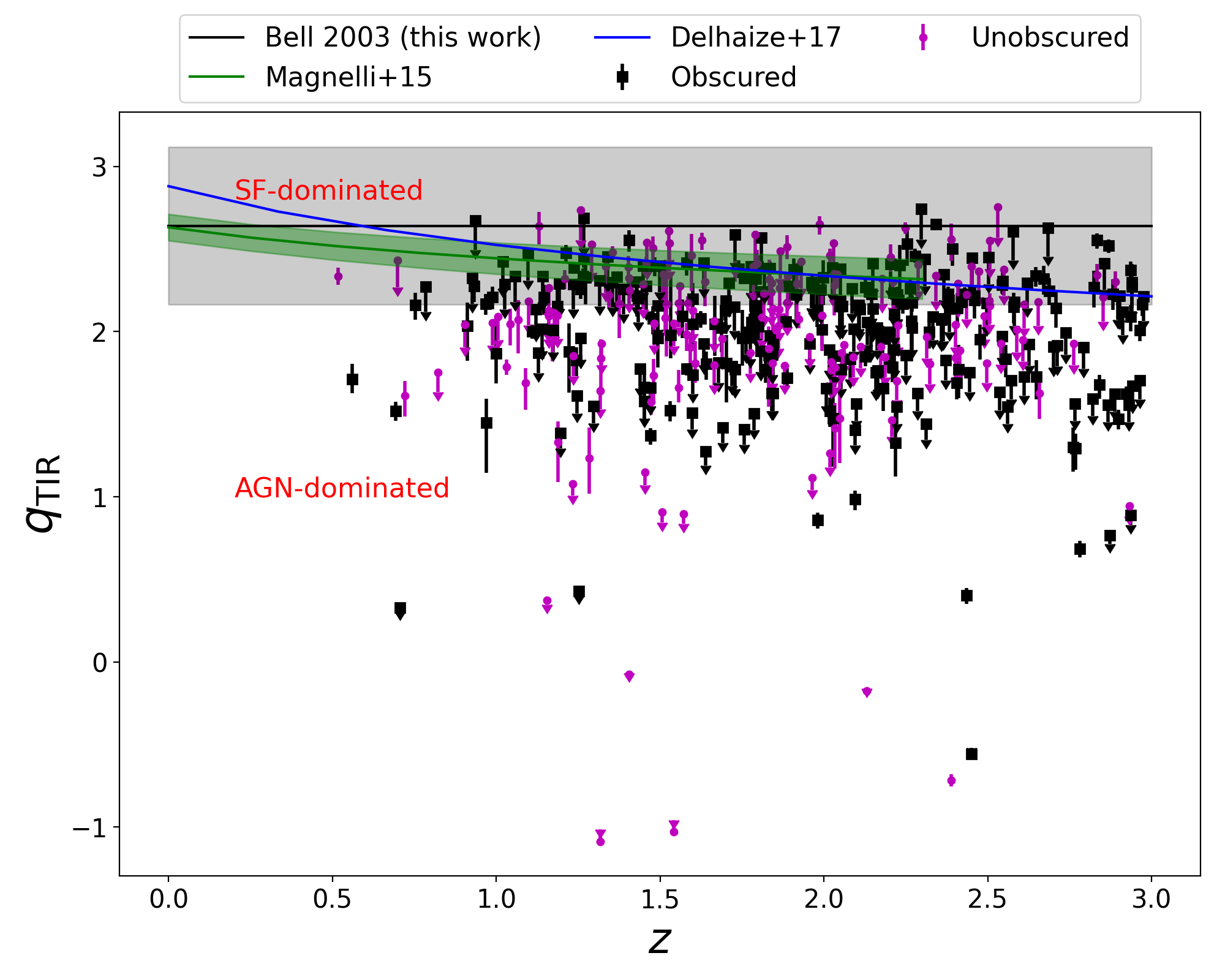}
\caption{Origin of the radio emission. {\it Left:} $8-1000 \rm \: \mu m$ SF luminosity ($L_{\rm SF, IR}$) versus $L_{\rm 1.4\: GHz}$. The right x-axis shows the SFRs associated with $L_{\rm SF, IR}$}. Black squares and magenta circles are obscured and unobscured IR quasars, respectively. In the case of sources with unconstrained $L_{\rm SF, IR}$, we plot the 99th percentile upper-limit of $L_{\rm SF, IR}$. Similarly, in the case of radio undetected sources, we plot 3$\sigma$ upper limits of $L_{\rm 1.4\: GHz}$. The continuous red line represents the region of the plane where all the observed $L_{\rm 1.4\: GHz}$ is expected to come from star-formation processes, assuming $q_{\rm TIR}=2.64$ \citep[][]{2003Bell}, and the dashed red lines illustrate a factor of three deviation from this line. {\it Right:} $q_{\rm TIR}$ versus $z$ for the radio detected quasar sample, together with the relations of \citet[][]{2015Magnelli} (green shaded region), \citet[][]{2017Delhaize} (blue line), and \citet[][]{2003Bell} (grey shaded area).  
\label{fig:L14_LSF}
\end{figure*}

We then investigate whether the excess in the radio emission of the obscured quasars comes from AGN or SF activity using the $q_{\rm TIR}$ parameter \citep[e.g.,][]{2003Bell,2015Magnelli, 2017Delhaize}, which quantifies the relative strength of the SF luminosity in the FIR band and the 1.4 GHz luminosity and is defined as


\begin{equation}
    { q_{\rm TIR}} = \log \left( \frac{L_{\rm SF, IR }}{3.75 \times 10^{12} \: L_{\rm 1.4 \: GHz}}  \right)
\end{equation}

\noindent where $L_{\rm SF, IR}$ is in units of W and $L_{\rm 1.4 \: GHz}$ in $\rm W \: Hz^{-1}$. As a first approach, we adopt $q_{\rm TIR, Bell}=2.64$ \citep[][]{2003Bell} and we consider that the radio emission is SF-dominated when  $ \log ( 1/3 \cdot 10^{q_{\rm TIR, Bell}}) < q_{\rm TIR}< \log ( 3 \cdot 10^{q_{\rm TIR, Bell}})$, following \citet[][]{2020Fawcett}. The left panel of Figure\,\ref{fig:L14_LSF} shows $L_{\rm SF, IR}$ against $L_{1.4 \rm \: GHz}$ and the dashed red lines indicate the region of the plane where we predict SF to dominate the radio emission. In the case of sources with an unconstrained $L_{\rm SF, IR}$, we use the 99th percentile of the $L_{\rm SF, IR}$ posterior distribution to decide if the radio emission is dominated by SF processes. We found that for $57\%$ (239/420) of the radio detected IR quasars, the radio emission is dominated by AGN processes. We also explore the origin of the radio emission in our sample using other relations for $q_{\rm TIR}$ reported in the literature. The right panel of Figure\,\ref{fig:L14_LSF} shows $q_{\rm TIR}$ against redshift for our sample, together with the relations of \citet[][]{2003Bell}, \citet[][]{2015Magnelli}, and \citet[][]{2017Delhaize}. The relationship we previously adopted (grey shaded area) is consistent with the other two relations over the explored redshift range. We note that we use $L_{\rm SF, IR}$ to calculate $q_{\rm TIR}$, while \citet[][]{2015Magnelli} and \citet[][]{2017Delhaize} use the total $8-1000 \rm \: \mu m$ luminosity; however, they carefully removed all the X-ray, IR, and radio AGNs from their samples, studying only the star-forming galaxies, where the total $8-1000 \rm \: \mu m$ luminosity is equivalent to $L_{\rm SF, IR}$ in our work. We stress that the calculated fractions of AGN-dominated radio quasars are lower limits since most quasars have an unconstrained $L_{\rm SF, IR}$, and in those cases, we use the 99th percentile of the $L_{\rm SF, IR}$ to compute $q_{\rm TIR}$.

We also compare the RL values of AGN- and SF-dominated radio sources initially assuming $q_{\rm TIR, Bell}=2.64$. For AGN-dominated radio sources, we find that the mean RL of obscured and unobscured quasars are $\mathcal{R}_{\rm AGN, obs}=-4.56\pm 0.03$ and $\mathcal{R}_{\rm AGN, unobs}=-4.65\pm 0.02$, respectively, and a comparison between the RL distributions gives an average p-value $\rm \approx 0.03$. For SF-dominated radio sources, the mean RL of obscured and unobscured quasars are $\mathcal{R}_{\rm SF, obs}=-4.86\pm 0.02$ and $\mathcal{R}_{\rm SF, unobs}=-5.10\pm 0.01$, with a mean $\rm p-value\approx 0.01$. These results suggest that the radio enhancement observed in obscured quasars likely comes from a mixture of SF and AGN processes and that the exact contribution from each process depends on the assumed relationship radio-SF luminosity. 


Table\,\ref{t:results} summarizes our results. Overall, we find strong evidence for obscured quasars having stronger radio emission than unobscured quasars, and we conclude that this radio enhancement comes from both SF and AGN processes. Interestingly, the enhancement of radio emission in SF-dominated radio sources provides some additional support for the result in Section\,\ref{sec:res:SFR} that obscured quasars have stronger SF luminosity (in FIR and radio) than the unobscured quasars.

\section{Discussion} \label{sec:discussion}

We have studied a complete sample of 578 IR quasars in the COSMOS field, with $L_{\rm AGN,IR}>10^{45}\rm \: erg\: s^{-1}$ and $ z \leq 3$. Our sample was carefully selected using detailed SED template fitting of the rest-frame 0.1-500 $\mu \rm m$ photometry. We guided the SED-fitting process by applying priors to the stellar mass using the observed relation of $M_{\rm BH}-M_{\rm \star}$ \citep[][]{2010Merloni} for sources with a $M_{\rm BH}$ measurement (otherwise, we use the mean $M_{\rm \star}$ of the obscured IR quasars), and applying a prior to the AD luminosity, using the observed relation of $L_{\rm 2500}-L_{\rm 2keV}$ \citep[][]{2016Lusso}. 
Our SED fitting results provide properties such as the AD reddening, the AD luminosity, the stellar masses of the host galaxies, IR luminosity from the AGN torus, and IR luminosity from the SF. We complement this information with the X-ray properties reported in the {\it Chandra COSMOS-Legacy} Survey, and the 1.4 and 3 GHz fluxes from \citet[][]{2017Smolcic} and \citet{2018Jin}, to investigate the strength of the radio emission from our IR quasars. We performed an extensive analysis of the AGN and host galaxy properties of our sample and compared the properties of the obscured and unobscured IR quasars, after matching both samples in redshift and rest-frame $6\rm \mu m$ AGN luminosity.

Overall, we found that more than 60\% of our IR quasar sample is obscured. For the X-ray bright sources, we distinguished between obscured and unobscured quasars by using the results from X-ray spectral fitting assuming an unobscured:obscured column density threshold of $N_{\rm H} = 10^{22} \rm \: cm^{-2}$. For the X-ray faint and undetected quasars, we used a combination of X-ray stacking and $L_{\rm 2-10keV}$--$L_{6\mu \rm m}$ analyses to demonstrate that the majority are heavily obscured, showing the power of our UV-to-FIR SED fitting to identify obscured quasars that are missed in the X-ray band.

On the basis of our detailed multi-wavelength and SED analyses, we found differences between the obscured and unobscured quasars which are not driven by luminosity and redshift. In Figure\,\ref{fig:composite_SEDs}, we showed that the median optical/UV-to-FIR SEDs of obscured and unobscured IR quasars are different -- the obscured quasars had a cooler AGN and stronger SF components than the unobscured quasars. These differences are driven by obscured quasars having a higher SF luminosity and {\it Herschel} detection fraction than unobscured quasars (see Table\,\ref{t:results}). 
Additionally, we found that the unobscured quasars have comparable stellar masses to the obscured quasars, implying that the stellar masses of the hosts are not responsible for the observed differences in the SF luminosity.

We also found robust evidence for excess radio emission from obscured quasars when compared to unobscured quasars (see Section\,\ref{sec:res:radio}). On the basis of comparisons between the radio and the IR emission from the AGN and the SF components, the radio emission appears to be dominated by AGN processes in at least 57\% of the radio-detected systems. The excess radio emission from the obscured quasars appears to be due to both SF and AGN processes, although the AGN contribution to the radio excess depends on the assumed relation for $q_{\rm TIR}$. 

In the following sections, we compare our results to previous studies and consider their implications. In Section\,\ref{sec:disc:sedshape}, we explore the origin of the differences in the AGN SED shape between the obscured and unobscured quasars. In Section\,\ref{sec:disc:radioSF}, we investigate the origin of the enhanced SF and radio emission from the obscured quasars. Finally, in Section\,\ref{sec:disc:evolution}, we discuss our results within the context of the AGN orientation and evolutionary models.

\subsection{Differences in the AGN SED shape between obscured and unobscured IR quasars} \label{sec:disc:sedshape}

\begin{figure*}
\centering

\includegraphics[scale=0.38]{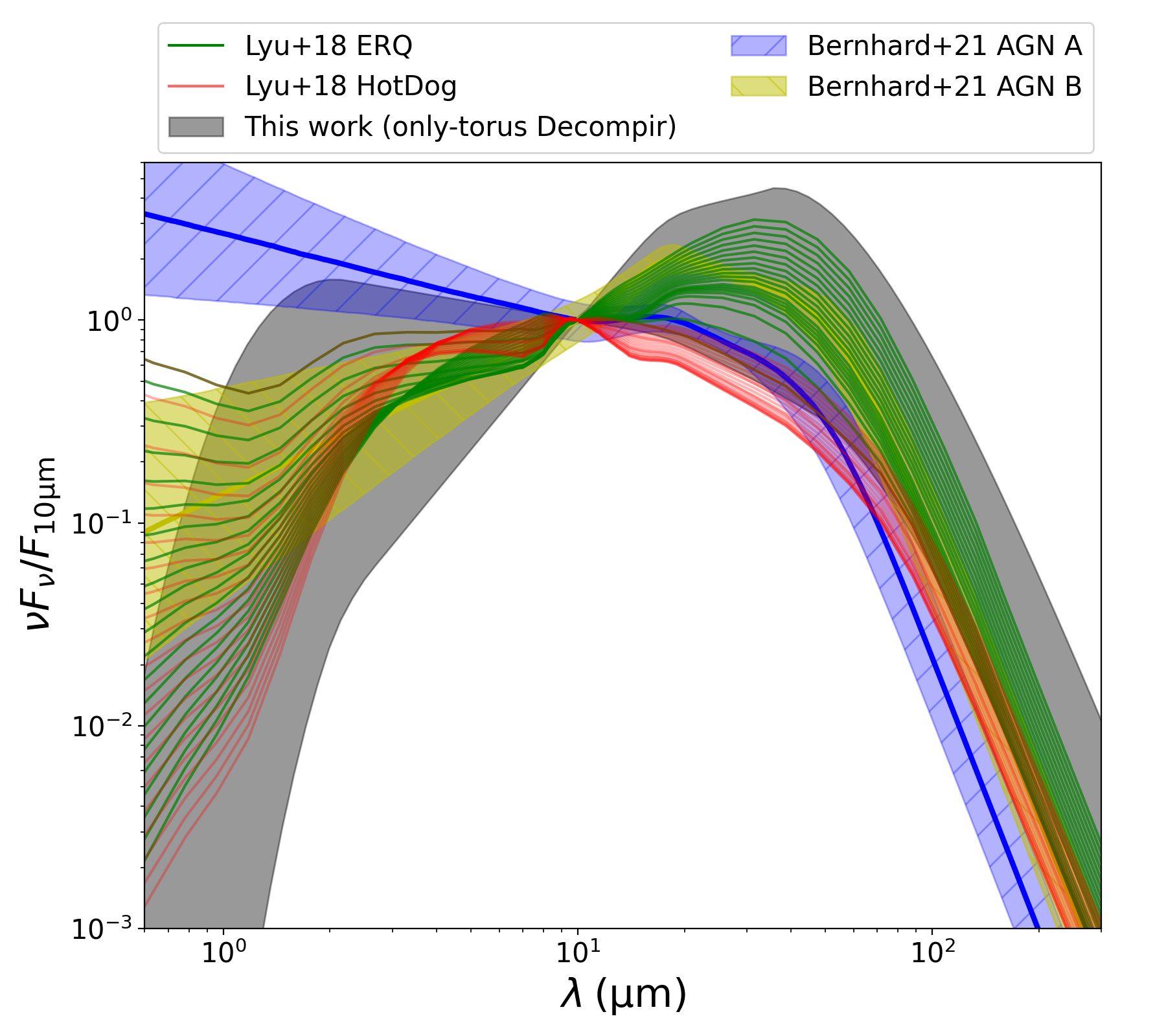}
\includegraphics[scale=0.38]{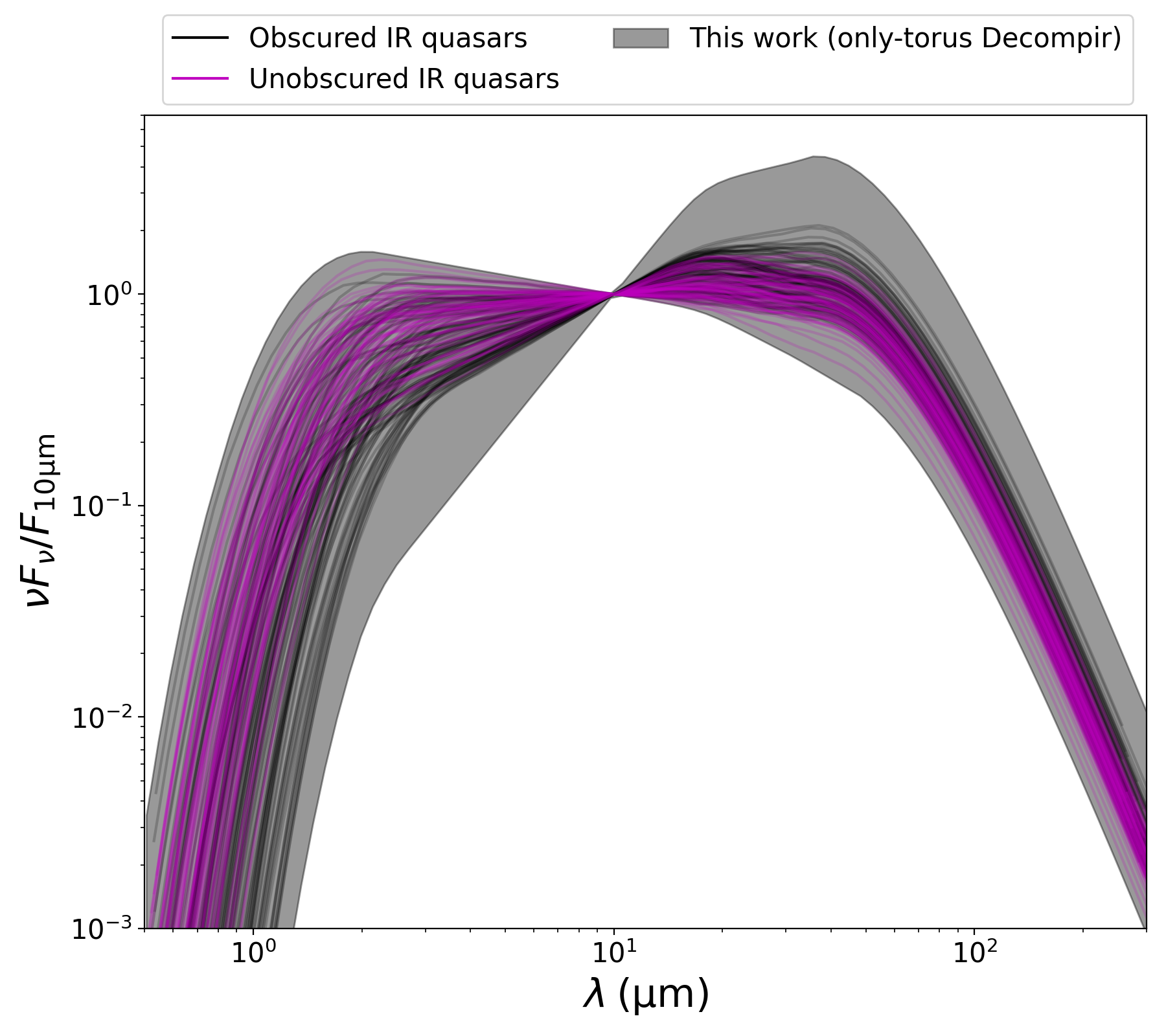}

\caption{{\it Left}: Comparison between the AGN model adopted in this work (DECOMPIR, \citealp[][]{2011Mullaney}), and the models constructed by \citet[][]{2021Bernhard} and \citet[][]{2018Lyu}. The grey shaded area represents the variety of shapes of DECOMPIR. The blue and yellow curves are the two AGN templates reported by  \citet[][]{2021Bernhard} and the blue and yellow shaded areas show the $1\sigma$ uncertainties of the templates. The light red lines represent the models for the hot dust-obscured galaxies of \citet[][]{2018Lyu} and the green lines are the models for extremely red quasars, in both cases for $\rm \tau_{ V}=[0,10]$. {\it Right}: best-fitting AGN models for unobscured (thin magenta lines) and obscured (thin black line) IR quasars together with the full variety of shapes allowed by DECOMPIR (grey shaded area). All the models are normalized by the flux at $10 \rm \: \mu m$.} 

\label{fig:CompModels}
\end{figure*}

Figure\,\ref{fig:composite_SEDs} shows that the median AGN SED shape of obscured quasars is cooler than that of unobscured quasars, with stronger emission at FIR wavelengths and weaker emission in the MIR band. These results are qualitatively consistent with previous studies that have demonstrated that Type-I AGNs have a hotter IR SED than Type-II AGNs \citep[e.g.,][]{2018Lyu,2021Bernhard}.


Recently, \citet[][]{2021Bernhard} studied the IR SED of 100 local AGNs from the {\it Swift}-BAT X-ray survey and showed that two different sets of AGN continuum templates are required to reproduce the diversity in AGN SEDs. They demonstrated that nuclear obscuration and an extended dusty structure in the polar direction are the main factors responsible for the observed SED differences. An extended dusty component in the polar direction has been found in a number of local (and mostly Type-II) AGN \citep[e.g.,][]{2012Honig,2014Tristram,2019Asmus}, and IR interferometry has shown that it can be extended over tens to hundreds of parsecs \citep[e.g.,][]{2016Asmus}. \citet[][]{2018Lyu} studied the impact of polar dust and nuclear obscuration on the IR SED of AGNs and demonstrated that the nuclear obscuration depletes the emission in the MIR band while the polar dust increases the emission in the FIR band ($\lambda \sim 30\: \mu m$, see Figure 19 of that paper). This polar dust emission is believed to be produced by AGN radiation pressure-driven outflows. Similarly, \citet[][]{2021CalistroG} found excess MIR emission for a statistical sample of red quasars (in comparison to blue quasars), again potentially associated with dusty outflows. Notably, the presence of energetic outflows could also help explain the possible radio enhancement due to AGN processes observed in obscured quasars, which we discuss in Section \ref{sec:disc:radioSF}.



An important question to ask is whether our results would significantly change if we used a different AGN model to fit the SEDs of our sources? Our implementation of the DECOMPIR model of \citet{2011Mullaney} allows for a wide variety of shapes, ranging from hot to cold SEDs. In the left panel of Figure\,\ref{fig:CompModels}, we show that our AGN model can broadly reproduce all the different SED shapes of the AGN models constructed by \citet[][]{2021Bernhard} and \citet[][]{2018Lyu}, and can even replicate very cold SEDs of extremely red quasars with a high polar dust optical depth ($\rm \tau_{ V}=\gtrsim 7-10$). We note that the deficit of emission at shorter wavelengths compared to \citet[][]{2021Bernhard} is due to that model including the emission from the AD (which is a different model component in our approach). We also compared our implementation of DECOMPIR with the models reported by \citet[][]{2006Richards}, \citet[][]{2016Symeonidis}, and \citet[][]{2020XuSun}, and found that it covers the SED shapes from all of these studies. We therefore conclude that our results would not significantly change if we adopted a different model to fit the AGN IR component of our sample. The right panel of Figure\,\ref{fig:CompModels} depicts the best-fitting AGN SED models of our obscured and unobscured IR quasars together with the full possible range of SEDs of DECOMPIR. It can be seen that even though DECOMPIR can reproduce very cold AGN SEDs, such extreme shapes were not required to fit the SEDs of our IR quasar sample.

\subsection{Origin of the SFR and radio enhancement of obscured quasars} \label{sec:disc:radioSF}

In Section\,\ref{sec:res:SFR}, we showed that obscured IR quasars have, on average, higher {\it Herschel} detection fractions and $\approx$~3 times higher average SFRs than unobscured quasars. Similar results were reported by \citet{2015Chen}, who found a significantly higher fraction of obscured quasars detected by {\it Herschel} at $250\: \rm \mu m$ and, correspondingly, a  $\approx 2$ times higher $L_{\rm SF, FIR}$ than unobscured quasars. Their analyses were based on shallower {\it Herschel} data in the Bo$\ddot{\rm o}$tes field, using predominantly photometric redshifts for the obscured quasars, but the overall agreement is striking. 

Evidence for an enhancement of the SF activity in obscured quasars has also been reported by \citet[][]{2004Page} and \citet[][]{2005Stevens}, who found differences in the submillimetre emission between X-ray obscured quasars and X-ray unobscured quasars, which imply that the SFRs of obscured systems are, on average, about an order of magnitude higher than those of unobscured systems. However, we note that all of those systems had broad optical emission lines and so were less obscured than our obscured quasars. A number of other studies have also explored the incidence of the AGN obscuration on the SF properties of the host galaxies, and although some have found tentative evidence for higher SFRs in obscured quasars, none have identified clear trends. Furthermore, many of these studies were based on small samples ($<30-40$ sources, \citealp[e.g.,][]{2013Georgantopoulos,2015Lanzuisi,2016DelMoro}), or used sample selection and obscuration criteria that were significantly different to that adopted in our study \citep[e.g.,][]{2014Merloni}. The main difference between our study and those mentioned above is the detailed UV/optical-to-FIR SED fitting performed on a substantial number of sources.

The other novel result from our research is the discovery that obscured quasars have stronger radio emission ($\approx 1.1$ times more radio detections and a $\rm RL \approx 0.2 \: dex$ higher) than unobscured quasars, a result not driven by differences in AGN luminosity or redshift between our samples. We obtained a qualitatively similar result when comparing red (mildly obscured) and blue (unobscured) quasars \citep[][]{2019Klindt,2020Fawcett,2020Rosario,2021Rosario,fawcett2022}, where red quasars were found to have a significant enhancement in compact radio emission ($<2\rm \: kpc$), likely arising from AGN-driven jets or winds. However, this is the first time that a radio enhancement in the most obscured AGNs has been identified. This result is particularly intriguing because the radio emission is optically thin; therefore, we do not expect the AGN radio properties to change with obscuration. Given that most of our quasars are in the radio-quiet regime, any nuclear-enhanced radio emission from obscured quasars is likely to be produced by either small-scale jets or shocks due to AGN-driven outflows \citep[e.g.,][]{2014Zakamska, 2015Nims,2021Jarvis, 2022Petley}. As we can not easily resolve the small-scale structure of our sample, distinguishing between compact jets and outflows is challenging. A possible test would be to investigate the radio spectral curvature of the sample, as jetted sources constrained by the ISM, are expected to show a strong correlation between their spectral turnover frequencies and their source sizes \citep[][]{1997ODea}.

Compact jets and AGN driven-outflows are thought to be crucial feedback mechanisms during the quasar phase \citep[e.g.,][]{2011Tombesi,2012WagnerBicknell,2014Cicone,2019Jarvis,2017HarrisonR,2020Costa}. Jets can interact with the interstellar medium, increasing the local velocity and producing jet-driven shocks, which might prevent the formation of stars \citep[e.g.,][]{2018Mukherjee}, but they also can compress the gas and trigger star formation \citep[e.g.,][]{2012Ishibashi, 2014Zubovas, 2022Bessiere}. Similarly, AGN driven-outflows can efficiently suppress the star formation of the galaxy by blowing out and destroying the high-density star-forming gas \citep[e.g.,][]{2020Costa}. Energetic outflows are clearly prevalent in AGN, particularly within luminous systems like quasars \citep[e.g.,][]{2008Ganguly,2014Harrison}. For example, \citet{2011Tombesi} studied a large sample of 42 local radio-quiet AGNs and found that $\gtrsim 35\%$ of the sample present ultra-fast outflows (UFOs), with a mean velocity of $42000 \: \rm km \: s^{-1}$. Interestingly, $80\%$ of the AGNs that are also Fe K absorbers (a ubiquitous signature in the X-ray spectra of obscured AGN) showed the presence of UFOs with a mean $N_{\rm H}$ value of $\sim 10^{23} \rm \: cm^{-2}$. Further evidence is provided by \citet[][]{2018DiPompeo}, who explored the kinematics of ionized gas via the [O III] $\lambda$5007 in a sample of MIR selected AGNs and found evidence for stronger outflows in obscured systems. Hence, powerful outflows might be particularly prevalent in the obscured quasars, potentially also at mild levels of obscuration \citep[][]{2021CalistroG}. Therefore, the nuclear processes that could explain the radio enhancement in obscured quasars could also explain the suppressed star formation observed in the unobscured quasars. We can explore the prevalence of outflows in our sample with NIR spectroscopy (i.e., VLT/XSHOOTER), which will cover the rest-frame optical [OIII] emission line.

We note that to compare the properties of obscured and unobscured quasars, we matched the sample in $L_{\rm 6 \mu m}$, which is an indicator of the dusty torus luminosity. \citet[][]{2016Stalevski} argued that due to the anisotropic nature of the torus, the ratio between the torus luminosity and the AGN bolometric luminosity changes for different covering factors, with type II AGNs having higher bolometric luminosities than type I AGNs for the same torus luminosity. This torus anisotropy could potentially explain the excess nuclear radio emission in the obscured quasars if the radio AGN emission is connected to the bolometric AGN luminosity. To test this hypothesis, we compare again the radio properties of the obscured and unobscured quasars, but this time focusing on the X-ray bright quasars and matching in $L_{\rm 2-10keV}$: the X-ray emission is used since it provides an isotropic measure of the overall luminosity of the AGN. We find the same overall results, the X-ray obscured quasars have a $\approx 0.2 \rm \: dex$ higher RL value and $\approx 1.1$ times more radio detections than the X-ray unobscured quasars. Consequently, we conclude that the torus anisotropy is not the primary reason for the excess radio emission observed in the obscured quasars.

\subsection{AGN orientation versus evolutionary models}\label{sec:disc:evolution}

In a scenario where the only difference between obscured and unobscured quasars is the orientation of the AGN with respect to our line of sight, we would expect both populations to have the same overall properties, after correcting for the impact of obscuration on the observed properties. Nonetheless, we have recently found evidence against a simple AGN orientation model due to the discovery of fundamental differences in the radio properties between red and blue quasars, which cannot be simply explained by the orientation of the AGN torus \citep[][]{2019Klindt,2020Fawcett,2020Rosario,2021Rosario,fawcett2022}. This basic result is in qualitative agreement with the radio differences between obscured and unobscured quasars found in this paper. However, interestingly, no significant differences in the SF properties between the red and blue quasars were found \citep[][]{2021CalistroG}, at least on average, in contrast with our results of the SF properties between obscured and unobscured quasars. Our results therefore extend those of the red and blue quasars but suggest a disconnect between the SF luminosity of red quasars and obscured quasars. We consider these results in the context of both AGN orientation and evolutionary scenarios below.


In the standard AGN orientation model, if the obscured IR quasar sample is obscured by the torus, we would not expect their SFRs to be three times higher than their unobscured counterparts (the AGN orientation model would predict no significant host-galaxy differences between obscured and unobscured quasars, on average). However, this assumes that the obscuration is dominated by dust in the nuclear region. Conversely, if a dominant fraction of the obscuring medium is extended over host-galaxy scales then the presence or absence of obscuration would not provide any inferences on the orientation of the AGN -- under this scenario, the identification of larger SFRs in the obscured quasars could be largely a selection affect (i.e.,\ larger SFRs would give an increased chance of line-of-sight obscuration through a dust-obscured SF region towards the quasar). Interestingly, some studies have shown that galaxies with SFRs of a few $\rm 100s \: M_{\odot} \: yr^{-1}$ can have interstellar medium (ISM) column densities above $10^{22} \rm \: cm^{-2}$ \citep[e.g.,][]{2017Buchner}, and could in extreme cases potentially extend to the CT regime \citep[e.g.,][]{2019Circosta, Gilli22}, depending on the compactness of the star formation. 
For example, in the recent study of \citet[][]{Gilli22}, deep ALMA data was used to constrain the column density of the ISM in AGN host galaxies and star-forming galaxies up to $z \approx 6$. They showed that at $z\leq 3$ the typical level of ISM obscuration is $N_{\rm H}\approx 10^{22}-10^{23} \rm \: cm^{-2}$, less than the level of obscuration for the majority of our obscured quasar sample, although up-to CT levels of host-galaxy obscuration were inferred for some higher redshift ($z>3$) systems. These  results suggest that at least a component of the obscuration from our obscured quasar sample could be produced over galactic scales.

We can provide a first-order test of the host-galaxy obscuration hypothesis by analyzing the SF properties of the obscured X-ray bright and obscured X-ray faint/undetected IR quasars. The obscured X-ray bright sample is dominated by sources with $N_{\rm H}\approx 10^{22}-10^{23} \rm \: cm^{-2}$, while the X-ray faint/undetected sources are expected to be heavily obscured or CT. If the quasars are obscured by their ISM, and taking a simple scenario where both samples have similarly compact star formation and host galaxies, we would expect the X-ray faint/undetected IR quasars to have higher SFRs than the X-ray bright IR quasars. However, as shown in Table~\ref{t:results} (see also Appendix\,\ref{sec:res:obsXraynoXray}), we find no significant differences in the SF properties between these obscured quasar sub-samples. This suggests that the majority of the obscuration occurs on AGN scales rather than galactic scales; however, we note that additional data is required to directly constrain the ISM obscuring column density and draw more robust conclusions.

\begin{figure}
\centering
\includegraphics[scale=0.42]{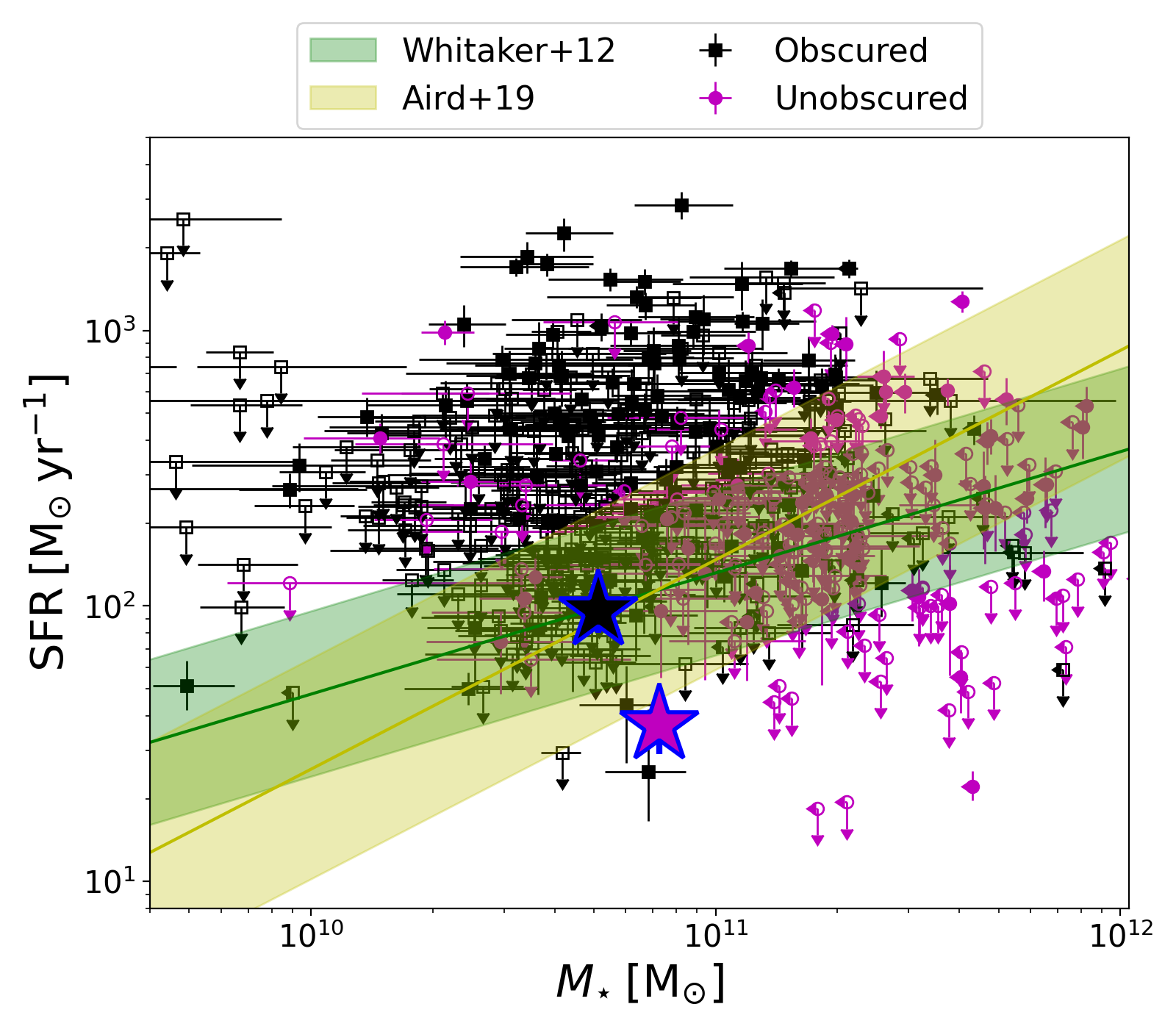}
\caption{Star-formation rate (SFR) against the stellar mass ($M_{\star}$) for obscured (black squares) and unobscured (magenta circles) IR quasars. Black and magenta stars with blue borders are the mean values of the matched obscured and unobscured IR quasar samples, respectively, calculated using \textsc{PosteriorStacker}. We also plot two relations for the main sequence of star formation from \citet[][]{2019Aird} (yellow shaded area) and \citet[][]{2012Whitaker} (green shaded area), assuming $\rm z=2$, the mean redshift of the sample. In the case of unconstrained sources, we plot 99th percentile limits for the SFR and $M_{\star}$ using open symbols.}
\label{fig:MS}
\end{figure}

If the obscuration of the obscured IR quasar sample occurs over AGN scales, our results would strongly disfavour the AGN orientation model, implying a more complex connection between obscured and unobscured quasars such as the quasar evolutionary model. Under this assumption we can use our results to test and characterize the evolutionary phases of obscured and unobscured IR quasars; for example, by comparing their SFRs with the main-sequence (MS) of star-forming galaxies, SFR--$M_{\rm \star}$ \citep[e.g.,][]{2004Brinchmann,2012Whitaker,2019Aird,2020Carraro}. Figure\,\ref{fig:MS} plots the SFR versus stellar mass of our quasar sample, together with two relations for the MS of star formation from \citet[][]{2012Whitaker} and \citet[][]{2019Aird}. We also show the mean SFR and stellar masses of the obscured and unobscured quasar sample calculated using \textsc{PosteriorStacker} (i.e., including upper limits). The mean SFR and $M_{\rm \star}$ of our sample indicate that obscured quasars have SFRs consistent with that of the MS, while unobscured quasars have more suppressed star formation (i.e., below the star-forming MS). This suggests that obscured quasars reside in galaxies growing more rapidly than those of unobscured quasars. These results are consistent with a scenario where obscured quasars are a young phase in the evolution of galaxies and quasars, where both the black hole and the galaxy are undergoing a period of rapid growth due to large reservoirs of dense gas and dust \citep[e.g.,][]{1988Sanders,2008aHopkins,2012Alexander,2014Lapi}, and are potentially hosting powerful winds \citep[e.g.,][]{2012Georgakakis}. Consequently, unobscured quasars would correspond to a later phase in galaxy evolution where the star formation is decreasing or is suppressed; for example, due to the ongoing impact of small-scale jets or AGN-driven outflows that either heat or drive out the high-density gas.

To put our results into context, recent simulations by \citet{2020Costa} showed that small-scale AGN-driven outflows could suppress the SF by a factor of $\approx 2-3$ over several tens of mega years (Myrs) timescale for AGN with $L_{\rm bol} \approx 10^{45}-10^{46}$~erg~s$^{-1}$ (equivalent luminosity range to our IR quasars) via the impact of energetic outflows with velocities of $\upsilon \approx 30000 \rm \:\: km\: \: s^{-1}$. These suppression timescales are reasonable for our sample, given that the typical AGN lifetime for sources with $\log \: L_{\rm AGN, IR}\approx 45.5$ (mean luminosity of our sample) is estimated to be of order $\sim 10 - 100 \rm \: Myr$ \citep[][]{2006Hopkins}. Consequently, the obscured quasars in our sample can plausibly represent an early evolutionary phase.  

Although this is the first time that clear empirical differences in the SFR and radio emission between obscured and unobscured quasars has been identified, tentative evidence that obscured quasars are younger systems than unobscured quasars have been previously reported. Several studies have shown that obscured quasars or AGNs are clustered more strongly than their unobscured counterparts (\citealp[e.g.,][]{2011Hickox,2014DiPompeo,2014Donoso,2018Powell}; Petter et al. in prep), where the SMBH is under massive relative to its host halo \citep[][]{2012Alexander}. Again, these results would argue against an AGN orientation model and favour an evolutionary scenario where obscured quasars are a younger phase in the evolution of galaxies and quasars. 

\begin{figure}
\centering
\includegraphics[trim={3.5cm 0 10cm 0},clip,scale=0.9]{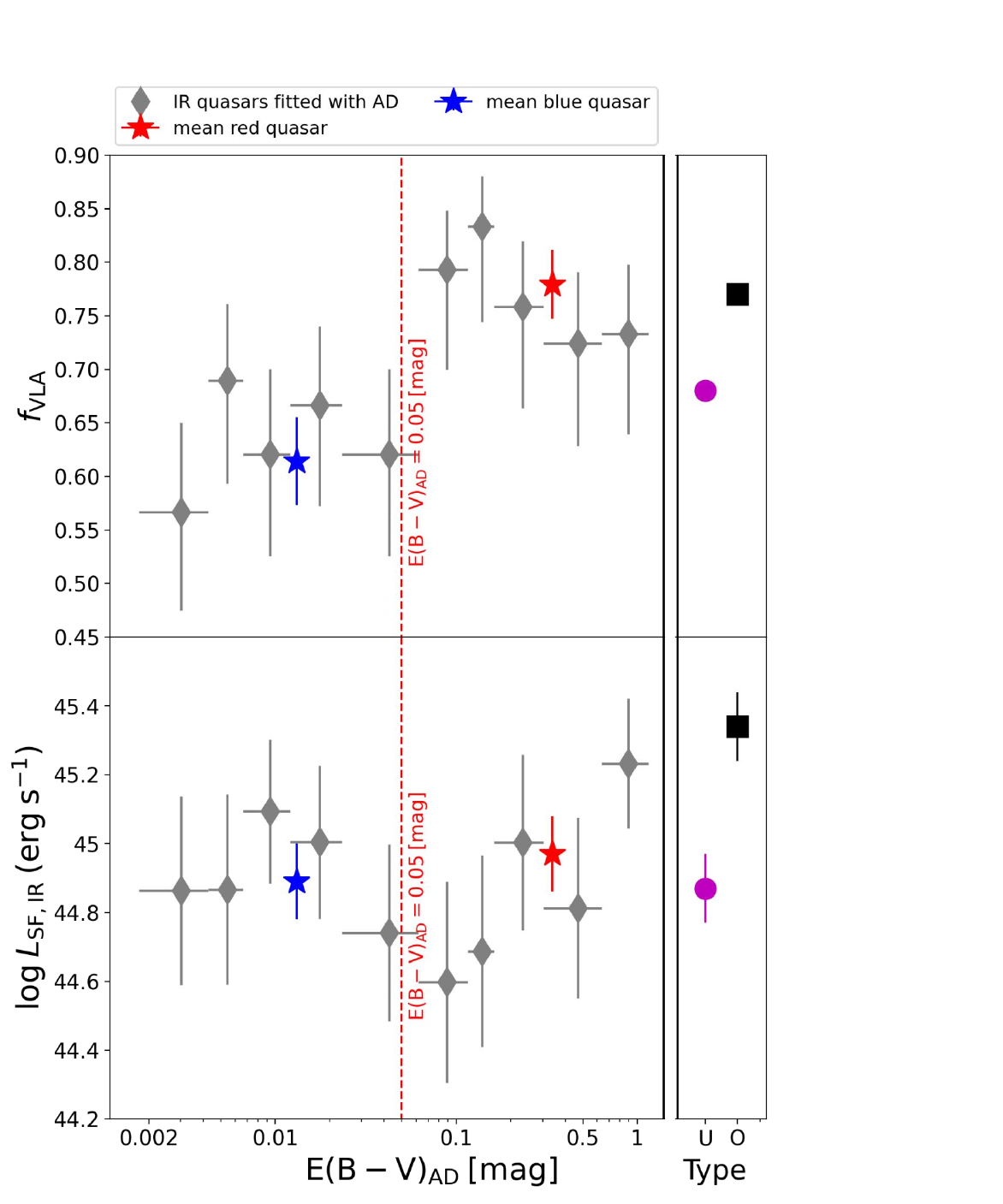}
\caption{Dependence of radio detection fraction and star-formation luminosity with obscuration. {\it Top panel}: Radio detection fraction ($f_{\rm VLA}$) versus the accretion disk reddening ($\rm E(B-V)_{AD}$) for sources fitted with an accretion disk component (grey diamonds, see Section\,\ref{sec:sedfitting} for details). The dotted red line indicates a reddening $\rm E(B-V)_{AD}=0.05 \: [mag]$, the blue/red quasars threshold. We also plot the mean $f_{\rm VLA}$ of blue quasars ($\rm E(B-V)_{AD}<0.05 \: [mag]$, blue star), red quasars ($\rm E(B-V)_{AD}\geq 0.05 \: [mag]$, red star). The right panel shows the values of unobscured quasars (magenta circle) and obscured quasars (black square), indicated as "Type". The error bars in the y-axes are binomial errors, and the x-axes are the size of the bins. {\it Bottom panel:} logarithmic $8-1000\rm \: \mu m$ SF luminosity ($L_{\rm SF,IR}$) versus the accretion disk reddening ($\rm E(B-V)_{AD}$) for sources fitted with an accretion disk component (grey diamonds, see Section\,\ref{sec:sedfitting} for details). The right panel shows the mean $L_{\rm SF,IR}$ for unobscured and obscured quasars. The error bars in the y-axes are 1$\sigma$ uncertainties, and the x-axes are the size of the bin. See the top-panel descriptions for the colour and symbol definitions.}
\label{fig:AD_RDF_LS}
\end{figure}


The prior red and blue quasar research from our group supported a scenario where red quasars are a younger phase in the evolution of galaxies and quasars than blue quasars \citep[][]{2019Klindt,2020Fawcett,2020Rosario,2021Rosario,fawcett2022}. To more directly connect our results to the red and blue quasars research, we analyzed the SF properties and radio-detection fraction of the IR quasars in our sample fitted with an AD component ($\approx$~80\% of the unobscured quasars; see Section\,\ref{sec:sedfitting} for details) to provide dust-reddening constraints towards the accretion disk. The top panel of Figure\,\ref{fig:AD_RDF_LS} shows the radio detection fraction versus obscuration. Following \citet[][]{fawcett2022}, we adopt $\rm E(B-V)_{AD}=0.05\:[mag]$ to distinguish between blue and red quasars. We can see that blue quasars (mean $f_{\rm VLA,blue}=0.61\pm0.04$) have consistently lower radio detection fractions than red quasars (mean $f_{\rm VLA,red}=0.78\pm0.03$). Interestingly, the radio detection fraction of the red quasars is practically the same as the obscured quasars, indicating that the observed differences between the radio detection fractions of unobscured and obscured quasars are predominantly due to the lower radio detection fractions of the blue quasars. The bottom panel of Figure\,\ref{fig:AD_RDF_LS} shows that the SF properties do not change with AD reddening and that blue and red quasars have consistent SF luminosities ($\log L_{\rm SF,IR,blue}/{\rm \: erg \: s^{-1}} = 44.89\pm0.11 $ and $\log L_{\rm SF,IR,red}/{\rm \: erg \: s^{-1}} = 44.97\pm0.11$, respectively), qualitatively consistent with the results of \citet[][]{2021CalistroG}; the red and blue quasars also have a mean SF luminosity consistent with unobscured quasars. These results provide further evidence that the radio excess of red quasars is due to AGN processes, as the SF luminosity remains constant, and that the radio excess does not keep increasing towards higher levels of obscuration. On the other hand, the consistency of the SF luminosity between red and blue quasars and the dramatic increase in SF luminosity towards more obscured systems suggests that red quasars may be an intermediate phase between blue quasars and obscured quasars where the SF has declined, but there is still sufficient opacity to cause enhanced radio emission from jets or winds shocking with the dense ISM or circumnuclear regions. Therefore, overall results are strongly consistent with the basic evolutionary sequence of obscured quasar--red quasar--blue quasar, characterized by a decrease in SF and opacity, likely driven by winds, outflows, shocks, and/or jets.

\section{Summary and Conclusions} \label{sec:conclusions}

We used the extensive multi-wavelength data in the COSMOS field to perform a detailed UV/optical-to-FIR SED fitting analysis of 578 IR quasars ($L_{\rm AGN,IR}>10^{45}\rm \: erg\: s^{-1}$) at redshift $ z\leq 3$. We carefully guided the SED fitting process by using priors based on observed relations to constrain the stellar mass and AD luminosity. We carried out an exhaustive analysis of the AGN and host galaxy properties of X-ray obscured and unobscured IR quasars using our SED fitting results together with the X-ray properties of our sample taken from the \textit{Chandra} COSMOS-Legacy survey, and the radio properties from \citet[][]{2017Smolcic} and \citet[][]{2018Jin}. Our results can be summarized as follows:

\begin{enumerate}

    
    \item Overall, 322 IR quasars ($56\%$ of the sample) are bright enough ($>30$ net counts) in the 0.5-7 keV band to have direct X-ray spectral fitting results and 99 are obscured by column densities $N_{\rm H}\geq 10^{22}\rm \: cm^{-2}$. The remaining 256 IR quasars are faint ($<30$ net counts in the 0.5-7 keV) or undetected in \textit{Chandra}. Using a combination of X-ray stacking, HR-z, and $L_{\rm 2-10 keV}$--$L_{6\rm \mu m}$ analyses, we demonstrate that the X-ray faint and undetected sources are heavily obscured by at least $N_{\rm H}\geq 10^{23}\rm \: cm^{-2}$ on average, and likely comprise the most obscured subset of the IR quasar population in the COSMOS field.  
    This result illustrates the power of IR SED fitting to find obscured AGN that would have been missed by even the deepest X-ray surveys (see Section\,\ref{sec:res:X-ray}): based on our sample $\approx$~61\% of the IR quasars are obscured ($N_{\rm H}\geq 10^{22}\rm \: cm^{-2}$).

    \item The median AGN SED shapes of obscured ($N_{\rm H}\geq 10^{22}\rm \: cm^{-2}$) and unobscured ($N_{\rm H}< 10^{22}\rm \: cm^{-2}$) quasars are different. Obscured quasars are cooler than unobscured quasars, having less emission at MIR wavelengths and stronger emission in the FIR waveband ($\lambda \gtrsim 30 \mu m$), potentially due to the presence of a extended polar component, beyond the torus (see Sections\,\ref{sec:res:SEDs} and\, \ref{sec:disc:sedshape} for a detailed analysis).
    
    \item We find an enhancement in the SFRs of obscured quasars when compared to unobscured quasars. Obscured quasars are $\approx 1.4(1.5)$ times more likely to have at least 1(2) {\it Herschel} detections, and their SFRs are $\approx 3$ times higher than those of unobscured quasars (see Sections\,\ref{sec:res:SFR} for details). We also find that the SF properties of blue ($\rm E(B-V)_{AD}<0.05 \: [mag]$) and red ($\rm E(B-V)_{AD}\geq 0.05 \: [mag]$ quasars are consistent, in qualitative agreement with our previous published research (see Section\,\ref{sec:disc:evolution}). 
    
    \item The mean stellar masses of unobscured quasars are broadly consistent with those of the obscured quasars ($M_{\rm \star, obs} = 10.71 \pm 0.02$ and $M_{\rm \star, unobs} = 10.86 \pm 0.02$). Consequently, the factor $\approx$~3 difference in SFR between the obscured and unobscured quasars cannot be driven by host-galaxy mass differences (see Section\,\ref{sec:res:Mstar}).
    
    \item We find 420 quasars (73\% of the sample) detected at VLA 1.4 or 3 GHz. Obscured quasars have higher radio detection fractions and radio loudness ($f_{\rm VLA}=77\pm 0.02 \%$ and $\mathcal{R}=-4.66 \pm 0.02$) than unobscured quasars ($f_{\rm VLA}=68 \pm 0.02\%$ and $\mathcal{R}=-4.85 \pm 0.01$), an enhancement which appears to have contributions from both AGN and SF processes (see Sections\,\ref{sec:res:radio} for details). We also find no significant difference in the mean radio detection fraction between red and obscured quasars, indicating
    that the strength of the radio emission does not significantly increase to higher levels of obscuration (see Section\,\ref{sec:disc:evolution}).
    
\end{enumerate}
    

In this work, we find strong evidence that obscured quasars have higher SFRs and excess radio emission in comparison to unobscured quasars. Our results disfavour a simple AGN orientation model but are in agreement with either extreme host galaxy obscuration in the obscured quasars or an evolutionary model where obscured quasars reside in an early phase of quasar evolution. Our analyses suggest that it is unlikely that host-galaxy obscuration can explain the extreme obscuration found for the majority of our obscured quasar sample, particularly for the X-ray faint and undetected systems. Consequently, we argue that obscured quasars are more likely to be an early phase of quasar evolution, characterized by higher SFRs and greater opacity, while red quasars are likely an intermediate phase between obscured and blue quasars, with suppressed SFRs but still a high opacity in the circumnuclear and/or interstellar medium. However, further observations are required to better constrain the relationship between obscured and unobscured quasars and to determine the origin and nature of the obscuration; for example, sub-millimetre observations to quantify the host galaxy obscuration and cold gas/dust content, UV--NIR spectroscopy to better disentangle the AGN from the SF emission and to search for the evidence of energetic outflows, and multi-frequency and high-resolution radio observations to identify the origin of the radio emission. 


\section*{Acknowledgements}

We thank the referee for his positive and constructive comments. We would also like to express our gratitude to Philip Hopkins for useful research discussions. We want to thank Mara Salvato for sharing her redshift compilation in the COSMOS field, and Stefano Marchesi and Giorgio Lanzuisi for sharing their X-ray spectral fitting results in COSMOS. This work has been supported by the EU H2020-MSCA-ITN-2019 Project 860744 “BiD4BESt: Big Data applications for black hole Evolution STudies.” DMA thanks the Science Technology Facilities Council (STFC) for support from the Durham consolidated grant (ST/T000244/1). CH acknowledges funding from an United Kingdom Research and Innovation grant (code: MR/V022830/1). AL is partly supported by the PRIN MIUR 2017 prot. 20173ML3WW 002 ‘Opening the ALMA window on the cosmic evolution of gas, stars, and massive black holes’. JP acknowledges support from STFC grants ST/T506047/1 and ST/V506643/1.

\section*{Data Availability}

The datasets generated and/or analysed in this study are available from the corresponding author on reasonable request.




\bibliographystyle{mnras}
\bibliography{references} 




\appendix

\section{IR quasars SED best-fitting parameters}

Table\,\ref{A:bfpar} report the best-fitting SED parameteres of our IR quasars.

\begin{landscape}

\begin{table}

\scriptsize
\scalebox{0.8}{
\begin{tabular}{rrrrlrrrrrrrrlrrrlrrrrrr}
 \hline
 \hline
 \noalign{\smallskip}

       ID &     Ra &  Dec &    z & ztype &  logLAGN &  logLAGN\_lo &  logLAGN\_up &  logL6um &   logLSF &  logLSF\_lo &  logLSF\_up &  logLSF\_99p &  uplim\_LSF &  logSM &  logSM\_lo &  logSM\_up &  uplim\_SM &  logL2500 &  logL2500\_lo &  logL2500\_up &  EBV &  EBV\_lo &  EBV\_up \\
       
    \noalign{\smallskip}  
        &     (deg) &  (deg) &    &  &  ($\rm erg/s$) &  ($\rm erg/s$) & ($\rm erg/s$) &  ($\rm erg/s$) &   ($\rm erg/s$) &  ($\rm erg/s$) & ($\rm erg/s$) &  ($\rm erg/s$) &   &  ($\rm M_{\odot}$) &  ($\rm M_{\odot}$) &  ($\rm M_{\odot}$) &  uplim\_SM &  ($\rm erg/s/Hz$) &  ($\rm erg/s/Hz$) &  ($\rm erg/s/Hz$) &  (mag) & (mag) & (mag) \\
       
    \noalign{\smallskip}  
     (1) &     (2) &  (3) &  (4) & (5) &  (6) &  (7) &  (8) &  (9) &   (10) &  (11) &  (12) &  (13) & (14) &  (15) &  (16) & (17) &  (18) &  (19) & (20) &  (21) &  (22) & (23) &  (24)\\

 \noalign{\smallskip}
\hline
\noalign{\smallskip}
  290049 & 149.54 & 1.73 & 2.20 & zphot &    45.12 &       44.94 &       45.27 & 44.59 &    43.78 &       42.67 &       44.88 &        45.83 &           True &        10.04 &            9.92 &           10.13 &              False &  27.29 &     26.47 &     28.13 &    0.16 &       0.04 &       0.45 \\
10124490 & 150.00 & 2.26 & 2.45 & zspec &    46.39 &       46.29 &       46.46 & 45.94 &    44.03 &       42.95 &       45.04 &        45.81 &           True &        11.11 &           11.03 &           11.18 &              False &   0.00 &      0.00 &      0.00 &    0.00 &       0.00 &       0.00 \\
  953566 & 150.56 & 2.75 & 2.77 & zphot &    46.26 &       46.13 &       46.36 & 45.54 &    43.94 &       42.73 &       45.13 &        45.90 &           True &        10.65 &           10.49 &           10.79 &              False &   0.00 &      0.00 &      0.00 &    0.00 &       0.00 &       0.00 \\
10217134 & 150.31 & 2.60 & 1.34 & zspec &    45.14 &       44.99 &       45.30 & 44.90 &    43.56 &       42.52 &       44.66 &        45.53 &           True &        10.71 &           10.41 &           10.99 &               True &  29.84 &     29.80 &     29.88 &    0.04 &       0.02 &       0.06 \\
  368580 & 150.07 & 1.85 & 1.21 & zphot &    45.34 &       45.08 &       45.54 & 44.84 &    45.96 &       45.90 &       46.01 &        46.07 &          False &        10.17 &            9.98 &           10.32 &              False &  29.50 &     29.38 &     29.61 &    0.66 &       0.58 &       0.73 \\
10126534 & 150.22 & 2.28 & 1.84 & zphot &    45.08 &       44.95 &       45.20 & 44.64 &    43.51 &       42.57 &       44.54 &        45.31 &           True &        10.46 &           10.30 &           10.67 &              False &  29.45 &     29.00 &     29.61 &    0.23 &       0.17 &       0.31 \\
10132783 & 149.93 & 2.36 & 1.10 & zspec &    45.70 &       45.40 &       45.83 & 45.13 &    44.80 &       43.07 &       45.53 &        45.72 &           True &        10.69 &           10.61 &           10.76 &              False &   0.00 &      0.00 &      0.00 &    0.00 &       0.00 &       0.00 \\
10110759 & 150.21 & 2.08 & 1.23 & zspec &    45.63 &       45.55 &       45.72 & 45.46 &    43.42 &       42.52 &       44.34 &        45.00 &           True &        10.50 &           10.21 &           10.77 &               True &  30.66 &     30.61 &     30.71 &    0.16 &       0.14 &       0.18 \\
  280625 & 150.29 & 1.71 & 2.29 & zphot &    45.34 &       45.17 &       45.53 & 44.90 &    44.86 &       43.18 &       45.47 &        45.77 &           True &        11.00 &           10.88 &           11.11 &              False &   0.00 &      0.00 &      0.00 &    0.00 &       0.00 &       0.00 \\
10261329 & 149.51 & 2.79 & 2.41 & zphot &    45.41 &       45.01 &       45.65 & 44.88 &    46.18 &       46.07 &       46.27 &        46.38 &          False &        10.62 &           10.51 &           10.77 &              False &   0.00 &      0.00 &      0.00 &    0.00 &       0.00 &       0.00 \\
\noalign{\smallskip}            
\hline
\end{tabular}}
\caption{IR quasars SED best-fitting parameters. Here we show the first 10 entries of the full table, which is available online. Col.(1): Object ID from \citet[][]{2018Jin}; Col.(2): right ascension from \citet[][]{2018Jin}; Col.(3) declination from \citet[][]{2018Jin}; Col.(4) redshift;  Col(5): redshift type; Col.(6): $8-1000\rm \: \mu m$ AGN luminosity; Col.(7): 16th percentile of the logarithmic $8-1000\rm \: \mu m$ AGN luminosity; Col.(8) 84th percentile of the logarithmic $8-1000\rm \: \mu m$ AGN luminosity; Col.(9): logarithmic $6\rm \: \mu m$ AGN luminosity; Col.(10): logarithmic $8-1000\rm \: \mu m$ star-formation luminosity ; Col.(11): 16th percentile of the logarithmic $8-1000\rm \: \mu m$ star-formation luminosity; Col.(12): 84th percentile of the logarithmic $8-1000\rm \: \mu m$ star-formation luminosity; Col.(13): 99th percentile of the logarithmic $8-1000\rm \: \mu m$ star-formation luminosity ; Col.(14): whether the star-formation luminosity is unconstrained; Col.(15): logarithmic stellar mass; Col.(16): 16th percentile of the logarithmic stellar mass; Col.(17): 84th percentile of the logarithmic stellar mass; Col.(18): whether the stellar mass is unconstrained; Col.(19): logarithmic accretion disk monochromatic luminosity at 2500\AA; Col.(20): 16th percentile of the logarithmic luminosity at 2500\AA; Col.(21): 84th percentile of the logarithmic luminosity at 2500\AA; Col.(22): accretion disk reddening; Col.(23):  16th percentile of the accretion disk reddening; Col.(24):  84th percentile of the accretion disk reddening. \label{A:bfpar}}

\end{table}

\end{landscape}

\section{Stacked X-ray images}

Here, we show the stacked images in the soft (0.5-2\,keV) and hard (2-8\,keV) bands of the X-ray faint ($<30$ net counts), X-ray undetected, and X-ray undetected with $SFR>100 \rm \: M_{\odot} \: yr^{-1}$ IR quasars. The three samples are significantly detected, suggesting the presence of AGN activity obscured by heavily-obscured column densities (see Section\,\ref{sec:res:X-ray}). 

\begin{figure*}

\centering
\includegraphics[scale=0.4]{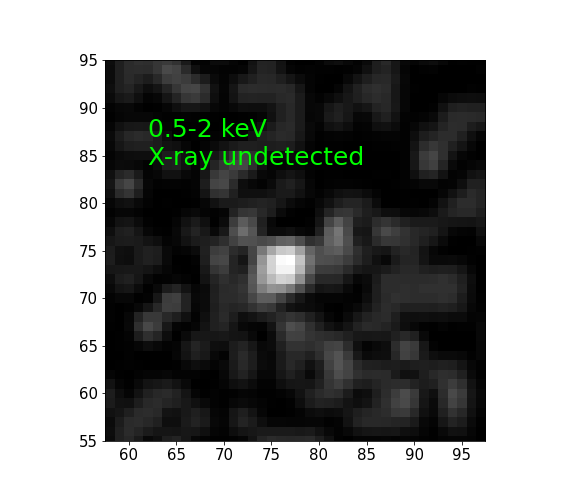}
\includegraphics[scale=0.4]{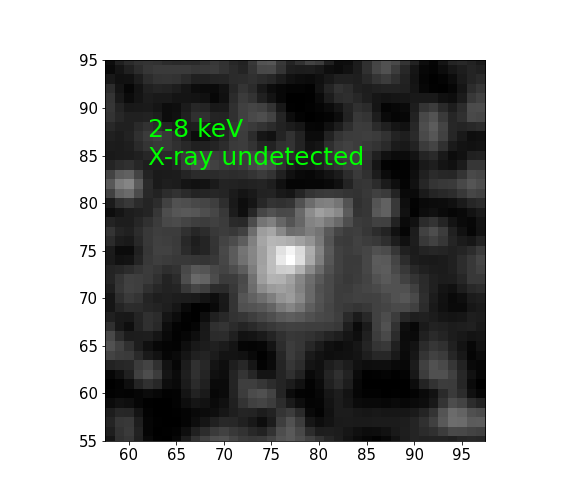}

\includegraphics[scale=0.4]{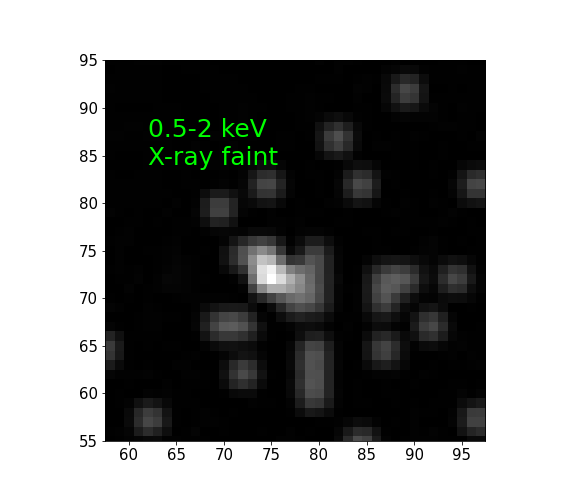}
\includegraphics[scale=0.4]{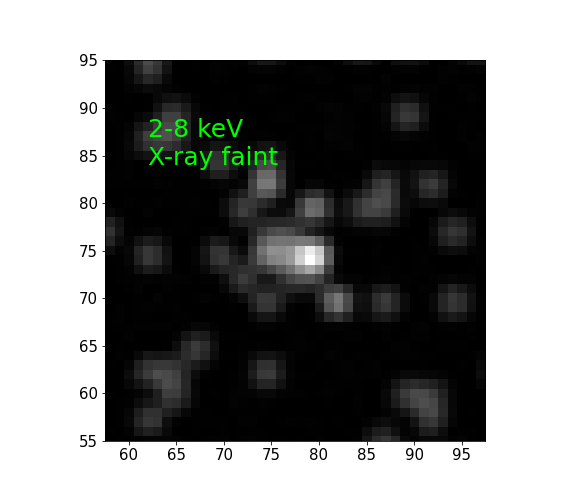}

\includegraphics[scale=0.4]{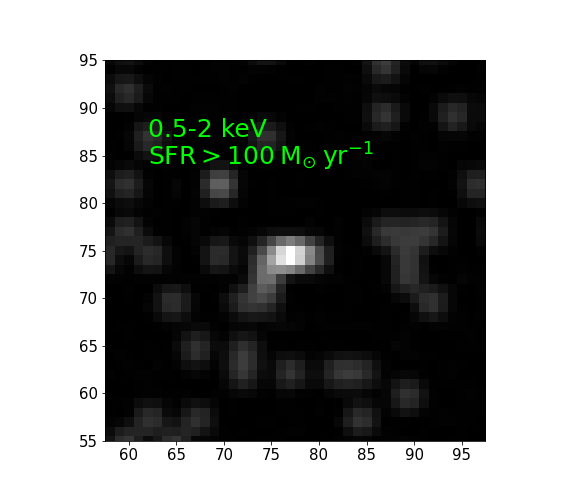}
\includegraphics[scale=0.4]{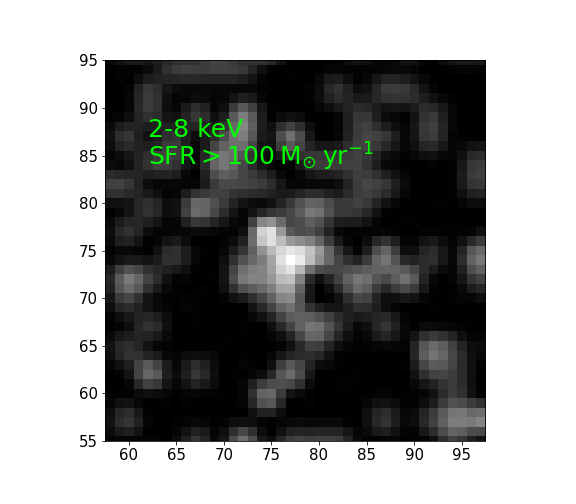}

\caption{X-ray stacked images in the 0.5-2 keV range (left plots) and 2-8 keV range (right plots) of X-ray undetected sources (upper plots), X-ray faint sources ($<30$ net counts, middle plots), and X-ray faint/undetected sources with a $\rm SFR>100\: M_{\odot} \: yr^{-1}$ calculated from the SED fits.   }
\label{fig:Xstack}
\end{figure*}

\section{X-ray detected and X-ray undetected obscured quasars}\label{sec:res:obsXraynoXray}

The majority of the obscured IR quasars are X-ray undetected or faint, and they lack individual absorbing column densities constraints. In order to provide additional evidence that shows that they are real AGN, in this section, we perform further analysis to that reported in Section\,\ref{sec:res:X-ray} (i.e., X-ray stacking and $L_{\rm 2-10\rm keV}$ -- $L_{\rm 6\: \mu m}$ analyses) to demonstrate that, on average, the X-ray faint and undetected quasars are heavily obscured and likely CT.

In Section\,\ref{sec:res:SFR}, we find that obscured quasars have SFRs three times higher than unobscured quasars, and that the majority ($72\%$) of the obscured sample is comprised of X-ray faint or undetected sources. To confirm that those sources with high SFR are real AGN and not just star-forming galaxies with red colours, we perform a similar analysis as that outlined in Section\,\ref{sec:res:X-ray} and stack the 62 positions of the X-ray undetected sources with a well-measured $L_{\rm SF, IR}$ and $\rm SFR>100 \: M_{\odot} \: yr^{-1}$. The results are reported in Table\,\ref{t:HR}. The bottom panels of Figure\,\ref{fig:Xstack} in the Appendix show the stacked images in the 0.5-2 keV and 2-8 keV bands. We obtain a hard X-ray signal, confirming that the sources are not intrinsically X-ray faint, but their X-ray emission is absorbed because they are heavily obscured. Following the same procedure of Section\,\ref{sec:res:X-ray}, we also estimate the obscuration of the sources through the HR and the $L_{2-10\rm keV}$--$L_{6\mu \rm m}$ relation. The right panel of Figure\,\ref{fig:obs_z} shows the position of the sources in the HR-$z$ plane, indicating that the X-ray undetected sources with high SFR are heavily obscured. This result is confirmed by Figure\,\ref{fig:Lx_L6um}, which shows the average positions of the sources in the $L_{2-10\rm keV}$--$L_{6\mu \rm m}$ plane, suggesting that the sources are affected by a Compton-thick column density. 


We then compare the AGN and host galaxy properties of X-ray detected and X-ray undetected/faint obscured IR quasars. If both samples are obscured AGN, we would expect them to have the same properties. The last two rows of Table\,\ref{t:results} show the mean properties of both samples after matching the obscured X-ray bright and X-ray faint/undetected quasars in redshift and $L_{\rm 6 \mu m}$. All the properties are broadly consistent within uncertainties, indicating that the only difference between them is that for X-ray faint and undetected quasars, most or all of the X-ray emission is absorbed by heavily obscured column densities.

\end{document}